\documentclass[useAMS,usenatbib]{mn2e}
\usepackage{graphicx,graphics}
\usepackage{amsmath}

\title[The survival of star clusters with BH subsystems]{The survival of star clusters with black hole subsystems}
\author[Long Wang]{Long Wang$^{1,2,3}$ \thanks{E-mail: longw@uni-bonn.de; }\\
$^{1}$ Argelander Institut F\"ur Astronomie, Auf Dem H\"ugel 71, 53121, Bonn, Germany \\
$^{2}$ Helmholtz-Institut f\"ur Strahlen- und Kernphysik, University of Bonn, Nussallee 14-16, D-53115 Bonn, Germany \\
$^{3}$ RIKEN Center for Computational Science, 7-1-26 Minatojima-minami-machi, Chuo-ku, Kobe, Hyogo 650-0047, Japan\\
}

\newcommand{\Eb}{|E_{\mathrm{2}}|}

\newcommand{\E}{|E|}
\newcommand{\Ekb}{E_{\mathrm{krh,2}}}
\newcommand{\Ekba}{E_{\mathrm{krh,2a}}}
\newcommand{\Ek}{E_{\mathrm{krh}}}

\newcommand{\Trhb}{T_{\mathrm{rh,2}}}
\newcommand{\Trhba}{T_{\mathrm{rh,2a}}}
\newcommand{\Trh}{T_{\mathrm{rh}}}
\newcommand{\Trhp}{T_{\mathrm{rh,p}}}
\newcommand{\Tcr}{T_{\mathrm{cr}}}
\newcommand{\Tcrs}{T_{\mathrm{cr,1}}}
\newcommand{\Trhs}{T_{\mathrm{rh,1}}}
\newcommand{\Tms}{T_{\mathrm{ms}}}

\newcommand{\mb}{m_{\mathrm{2}}}
\newcommand{\m}{\langle m \rangle}
\newcommand{\ms}{m_{\mathrm{1}}}

\newcommand{\Nb}{N_{\mathrm{2}}}

\newcommand{\N}{N}
\newcommand{\inttms}{T[\Tms]}

\newcommand{\inttrhp}{T[\Trhp]}

\newcommand{\vave}{\langle v \rangle}
\newcommand{\vb}{v_{\mathrm{2}}}
\newcommand{\rhb}{R_{\mathrm{h,2}}}

\newcommand{\rh}{R_{\mathrm{h}}}
\newcommand{\rhs}{R_{\mathrm{h,1}}}
\newcommand{\rt}{R_{\mathrm{t}}}
\newcommand{\resc}{R_{\mathrm{esc}}}
\newcommand{\rtid}{R_{\mathrm{tid}}}
\newcommand{\M}{M}
\newcommand{\Mb}{M_{\mathrm{2}}}
\newcommand{\MbMave}{\langle \Mb/\M \rangle}
\newcommand{\Mbdot}{\dot{M}_{\mathrm{2}}}
\newcommand{\Ms}{M_{\mathrm{1}}}
\newcommand{\comb}{\ln \Lambda}
\newcommand{\combb}{\ln \Lambda_{\mathrm{2}}}

\newcommand{\vp}{\langle (\Delta v_{||})^2 \rangle}

\newcommand{\nk}{n_{\mathrm{k}}}
\newcommand{\ns}{n_{\mathrm{1}}}
\newcommand{\n}{\langle n \rangle}
\newcommand{\va}{\langle v \rangle}
\newcommand{\nb}{n_{\mathrm{2}}}
\newcommand{\mk}{m_{\mathrm{k}}}
\newcommand{\vk}{v_{\mathrm{k}}}
\newcommand{\vs}{v_{\mathrm{1}}}
\newcommand{\ka}{k_{\mathrm{1}}}
\newcommand{\kap}{k^{\prime}_{\mathrm{1}}}
\newcommand{\kp}{k^{\prime}}
\newcommand{\kb}{k_{\mathrm{2}}}
\newcommand{\psib}{\psi_{\mathrm{2}}}
\newcommand{\Tscale}{T_{\mathrm{scale}}}
\newcommand{\tst}{t_{\mathrm{0}}}
\newcommand{\vej}{v_{\mathrm{ej}}}
\newcommand{\vejm}{v_{\mathrm{ej,m}}}
\newcommand{\vescm}{\langle v_{\mathrm{esc}} \rangle}
\newcommand{\vesc}{v_{\mathrm{esc,c}}}
\newcommand{\ahs}{a_{\mathrm{hs}}}
\newcommand{\Tdis}{T_{\mathrm{dis}}}
\newcommand{\Tdisp}{T_{\mathrm{dis,p}}}
\newcommand{\Tdisps}{T_{\mathrm{dis,p1}}}
\newcommand{\aimf}{\alpha_{\mathrm{3}}}
\newcommand{\Mse}{M_{se}}
\newcommand{\mbave}{\langle \mb \rangle}
\newcommand{\msave}{\langle \ms \rangle}
\newcommand{\mbmtr}{\mathcal{R}_{\mathrm{tr}}}
\newcommand{\gb}{\gamma_{\mathrm 2}}

\date{Accepted --. Received --; in original form --}

\pagerange{\pageref{firstpage}--\pageref{lastpage}} \pubyear{2019}

\begin{document}

\maketitle

\label{firstpage}

\begin{abstract}
  Recent observations have detected top-heavy IMFs in dense star forming regions like the Arches cluster.
  Whether such IMFs also exist in old dense stellar systems like globular clusters is difficult to constrain, because massive stars already became black holes (BHs) and neutron stars (NSs).
  However, studies of stellar dynamics find that BHs/NSs influence the long-term evolution of star clusters.
  Following \cite{Breen2013} and by carrying out two-component $N$-body simulations, we demonstrate how this dynamical impact connects with the shape of IMFs.
  By investigating the energy balance between the BH subsystem and the global, we find that to properly describe the evolution of clusters, a corrected two-body relaxation time, $\Trhp=\Trh/\psi$, is necessary.
  Because $\psi$ depends on the total mass fraction of BHs, $\Mb/\M$, and the mass ratio, $\mb/\ms$, the cluster dissolution time is sensitive to the property of BHs or IMFs.
  Especially, the escape rate of BHs via ejections from few-body encounters are linked to mass segregation.
  In strong tidal fields, top-heavy IMFs easily lead to the fast dissolution of star clusters and the formation of BH-dominant dark clusters, which suggests that the observed massive GCs with dense cores are unlikely to have extreme top-heavy IMFs.
  With the future observations of gravitational waves providing unique information of BHs/NSs, it is possible to combine the multi-message observations to have better constrains on the IMFs of old star clusters.

\end{abstract}


\section{Introduction}
\label{sec:introduction}

Whether the initial mass function (IMF) is environmentally dependent is a long existing question.
The observations of present-day star forming regions (young star clusters and associations) in the Galaxy and Magellanic clouds show no strong evidence that the high-mass end of IMFs ($>1 M_\odot$) depends on the density for the range of $\rho_{\mathrm 0} = 35-3\times10^4 M_\odot \mathrm{pc}^{-3}$, where $\rho_{\mathrm 0}$ is the central density of clusters, and the metallicity $Z = 0.2-1 Z_\odot$ \citep{Bastian2010}.
However, extremely dense environments like the Arches cluster ($\rho \approx 2\times 10^5 M_\odot \mathrm{pc}^{-3}$), the Galactic center and 30~Doradus are found to contain top-heavy IMFs with heavy-end exponent $\alpha \approx -1.7 \sim -1.9$ \citep{Lu2013,Schneider2018,Hosek2019}.
Recently, by measuring the ${}^{13}$CO/C$^{18}$O ratio in starburst galaxies at high redshift ($z\sim 2-3$), \cite{Zhang2018} found that a top-heavy IMF with $\alpha \approx -2.1$ is necessary to explain the star formation rate there.
Thus, there is a possible trend that IMF might be top-heavy in high-density environments.
If this is the case, an interesting question is whether old and dense globular clusters (GCs) in the Galaxy also contain top-heavy IMFs.
It is difficult to determine IMFs in GCs since all massive stars have already evolved to black holes (BHs) and neutron stars (NSs) a long time ago.
A semi-analytic work by \cite{Marks2012} estimated the IMFs of GCs and suggested that GCs can contain extreme top-heavy IMFs ($\alpha \approx -0.7 \sim -2.3$) depending on density and metallicity.
However, a recent study by \cite{Baumgardt2017} compared the scaled $N$-body models with observational data of $35$ GCs found that the present-day mass function of GCs agrees well with \cite{Kroupa2001,Chabrier2003} IMFs.
Thus, the studies of IMFs in GCs are controversial and there is no star-by-star $N$-body simulations of GCs yet to clarify this due to the time-consuming computing.

One significant influence of IMFs (heavy-mass end) is the number fraction of BHs and NSs formed in star clusters.
By including the BH subsystems in fast Monte-Carlo simulations, \cite{Chatterjee2017} and \cite{Giersz2019} found that tidal filling GCs with top-heavy IMFs dissolve much faster than the models with Kroupa/Chabrier IMFs, thus more difficult to survive until today.
Therefore, the dynamical effect of BHs may provide a strong constraint on the IMFs of star clusters, especially in GCs.

It is very useful to develop a theory how BHs dynamically affect the evolution of star clusters, in order to provide a fast method to describe the properties of BHs and star clusters without expensive $N$-body models.
\cite{Henon1961,Henon1975} showed that the long-term evolution of star clusters are guided by the balance of energy flux, i.e. the energy generation in the cluster core is regulated by the energy demand of the system.
Based on this idea, \cite{Breen2013} suggested that in star clusters with many BHs, the energy generation is dominated by the few-body encounters between BH binaries and singles in a centrally concentrated BH subsystem.
Thus, the balance of energy flux can be described by \citep[Eq.~1 in ][]{Breen2013}
\begin{equation}
  \frac{\E}{\Trh} \approx k \frac{\Eb}{\Trhb} ,
  \label{eq:eb}
\end{equation}
where $\Eb$ and $\E$ are the total energy of the BH subsystem and the global system respectively; $\Trh$ and $\Trhb$ are the two-body relaxation times measured at the half-mass radius of the system, $\rh$, and the BH subsystem, $\rhb$, respectively.
Hereafter, the suffix ``2'' and ``1'' represent BH subsystems and light star components separately.
When the global parameters of a cluster are known, this relation provides a constrain on the properties of BH subsystems.
It results in several important features of the long-term evolution of star clusters:
\begin{itemize}
\item During the balanced stage, $\rh$ increases due to the heating of BHs and $\rhb/\rh$ has a certain relation \citep[Eq.~4 in][]{Breen2013}.
\item BHs are trapped in the deep potential of the cluster, thus the escape of BHs via tidal stripping can be ignored. The ejection of BHs via few-body encounters between BH binaries and singles in the cluster core is the major channel to remove BHs from the cluster. Since the energy generation rate via the few-body encounters is controlled by the global energy requirement, the escape rate of BHs is independent of their relaxation time, $\Trhb$, but scales with $\Trh$ of the global system \citep[Eq.~9 in ][]{Breen2013}.
\item In tidally limited systems, if the total mass fraction of BHs is small, i.e. $\Mb/\M<0.1$, the BHs are depleted faster and $\Mb/\M$ decreases as clusters evolve. Otherwise $\Mb/\M$ increases and finally ``dark clusters'' form \citep{Banerjee2011,Giersz2019}.
\end{itemize}
The first two points assume that BHs are not the dominant component in star clusters, which is the general case for canonical IMFs \citep{Kroupa2001,Chabrier2003}.

On the other hand, with the help of $N$-body simulations, \cite{Baumgardt2001,Baumgardt2003} developed a theory for the mass loss of star clusters and showed that the dissolution time of the systems is \citep[by putting Eq.~1 to Eq.~5 from][and canceling $R_{\mathrm G}/V_{\mathrm G}$ in Eq.~5]{Baumgardt2003}
\begin{equation}
  \Tdis \sim \Trh^x\Tcr^{1-x} (\rt/\rh)^{3/2} ,
  \label{eq:tdis}
\end{equation}
where $\Tcr$ is the crossing time, $\rt$ is the tidal radius of the system.
The power index $x$ depends on the structure of the cluster and the tidal field, and is around $0.75$.
The term $\Tcr^{1-x}$ represents the back-scatter effect of potential escapers and the term $(\rt/\rh)^{3/2}$ is a  radius scaling factor that scales the half-mass radius by the tidal radius.
This theory provides a simple solution to evolve the masses of star clusters by given initial conditions.
However, the effect of BH subsystems is not included in their study, thus it is unknown whether the theory can be extended to clusters with top-heavy IMFs.

In this work, as a follow-up study of \cite{Breen2013}, we extend the energy-balance theory to the region of high $\Mb/M$, which represents star clusters with top-heavy IMFs.
In Section~\ref{sec:topimf}, how top-heavy IMFs affect $\Mb/M$ is shown.
Then we describe the initial conditions of the two-component $N$-body models in Section~\ref{sec:method}.
We provide the results and the analysis about the energy-flux balance, the mass loss of BHs and stars in isolated environments in Section~\ref{sec:isolated} and in the tidally filling case in Section~\ref{sec:tidal}.
Finally, we discuss our results and draw conclusions in Section~\ref{sec:conclusion}.

\section{Top-heavy IMF}
\label{sec:topimf}

The top-heavy IMFs can be defined by varying the slope, $\aimf$, in the heavy end of IMFs ($1-150 M_\odot$) based on the multi-component power-law IMF \citep[][]{Kroupa2001},
\begin{equation}
  \begin{aligned}
    &&\xi(m) \propto m^{\alpha_{\mathrm i}} & \\
    \alpha_{\mathrm{1}} & = -1.3,&   0.08 \le m/M_\odot &< 0.50 \\
    \alpha_{\mathrm{2}} & = -2.3,&   0.50 \le m/M_\odot &< 1.00 \\
    \alpha_{\mathrm{3}} &,       &   1.00 \le m/M_\odot &< 150.0.\\ 
  \end{aligned}
  \label{eq:imf}
\end{equation}
The total mass fraction of BHs, $\Mb/M$, after $100$~Myr evolution can be estimated by using the updated \textsc{sse/bse} code \citep{Hurley2000,Banerjee2019}.
The models of stellar winds and formation of NSs and BHs are detailed discussed in \cite{Banerjee2019}.
The remnant mass is based on the rapid supernovae scenario from \cite{Fryer2012} and the calculation also includes the pair-instability supernovae \citep{Belczynski2016}.
The result is shown in Table~\ref{tab:imf}.
Here all formed BHs are counted in the calculation and the metallicity of stars is $Z=0.001$.
A part of BHs may immediately escape the clusters due to high kick velocities after the supernovae.
This effect is ignored in our analysis.
For the case of a canonical IMF of \cite{Kroupa2001} with $\aimf=-2.3$, after $100$~Myr, BHs contribute about $7\%$ of the total mass and about $20\%$ of the initial mass is lost via stellar evolution (winds) and supernovae.
When $\aimf=-1.7$, which is the case for the Arches cluster \citep{Hosek2019} and the Galactic center \citep{Lu2013}, the mass fraction of BHs grows significantly (to $38\%$).
The average mass ratio for BHs and stars, $\mbave/\msave$, is about $40$ for metallicity $Z=0.001$.
This result can be altered (not significantly) by choosing different assumptions in the stellar evolution model.
But in general the shape of the IMF changes $\Mb/\M$ strongly.

\begin{table}
  \centering
  \caption{The BH properties after $100$~Myr stellar evolution for different IMFs with the metallicity, $Z=0.001$. Columns show the mass fraction of BHs, $\Mb/M$, the stellar wind mass loss normalized to the initial total mass, $\Mse/M(0)$, the number fraction of BHs, $\Nb/N$, and the average mass ratio for BHs and stars, $\mbave/\msave$.}
  \label{tab:imf}
  \begin{tabular}{lllll}
    \hline
     $\aimf$    & $\Mb/M$ & $\Mse/M(0)$ & $\Nb/N$ & $\mbave/\msave$ \\
    -1.5 & 0.554   & 0.536       & 0.0319  & 37.8      \\
    -1.7 & 0.376   & 0.466       & 0.0151  & 39.3      \\
    -2.0 & 0.182   & 0.334       & 0.00522 & 42.5      \\
    -2.3 & 0.0733  & 0.204       & 0.00181 & 43.7      \\
    \hline
  \end{tabular}
\end{table}

\section{N-body models}
\label{sec:method}

A BH is much more massive than a light star.
It is expected that the stellar dynamical effect of BHs play an important role for the long-term evolution of star clusters.
With top-heavy IMFs, this effect is more pronounced.
In order to investigate this, we carry out a series of $N$-body simulations of star clusters with two components and no stellar evolution \citep[similar to][]{Breen2013}.
The spherically symmetric Plummer models are employed for initial conditions.
Two components are homogenously mixed initially with the same density and velocity profile.
Each component has an equal mass for all members.
The heavy and light components represent BHs and light stars separately.
Several physical processes involved in the evolution of a real star cluster are ignored in these simple models.
However, in the first step, it is important to obtain a better theoretical understanding by investigating the simplest model with the interesting physical process well isolated.

The initial number fraction of the heavy component, $\Nb/N(0)$, and the average stellar mass ratio, $\mb/\ms$, are varied for different models.
For convenience, hereafter ``m[value]'' and ``N[value]'' represent different $\mb/\ms$ and $\Nb/N(0)$ separately (e.g. m2.5 is $\mb/\ms=2.5$ and N0.01 is $\Nb/N(0)=0.01$).
The initial total mass ratio of two components, $\Mb/M(0)$, are listed in Table~\ref{tab:init}.
$\Mb/M(0)$ vary from $2.5\%$ to $67\%$, thus this model set represents a wide range of top-heavy IMF.
$\mb/\ms$ includes the values of $2.5$ to $40$, corresponding to the masses of NSs and BHs.
The initial total number of objects have two values, $32k$ and $64k$, as shown in Table~\ref{tab:init}.

For each model, we consider the star clusters in an isolated environment and in a point-mass potential with a circular orbit.
A star is defined as an escaper if its energy is positive and its distance to the cluster center is larger than $\resc$.
In the isolated case, $\resc = 20~\rh$.
In the tidal filling case, $\resc = 2~\rtid$, and initially, $\rtid/\rh\approx7.3$, for all models, which is close to the case of the tidal filling King model with $W=6\sim7$.

The \textsc{nbody6++gpu} code \citep{Wang2015} is used to perform the $N$-body simulations.
The code is designed to simulate star clusters with regularization algorithms to accurately deal with the strong interaction between stars and BHs, and therefore fits the purpose of this study \citep[][and reference therein]{Aarseth2003}.

\begin{table}
  \centering
  \caption{The initial $\Mb/M(0)$ for all two-component $N$-body models.
    Rows show the same $\mb/\ms$ and columns show the same $\Nb/N(0)$.
    The symbol ``*'' indicates that the model has initial $N(0)=64k$ and ``**'' indicates that both $N(0)=32k$ and $N(0)=64k$ exist. All other models have initial $N(0)=32k$.
  }
  \begin{tabular}{lllrr}
    \hline
    $\frac{\mb}{\ms} \backslash \frac{\Nb}{\N}(0)$ & 0.005     & 0.01         & 0.03  & 0.05  \\
    2.5 &           & 0.025        & 0.072 & 0.116 \\
    5   &           & 0.048        & 0.134 & 0.208 \\
    10  & $0.048^{*}$ & $0.092^{**}$ & 0.236 & 0.345 \\
    20  &           & $0.168^{**}$ & 0.382 & 0.513 \\
    40  &           & $0.288^{**}$ & 0.553 & 0.678 \\
    \hline
  \end{tabular}  
  \label{tab:init}
\end{table}

\section{Isolated star clusters}
\label{sec:isolated}
\subsection{Energy flux}
\label{sec:energy}

\begin{figure}
  \centering
  \includegraphics[width=0.9\columnwidth]{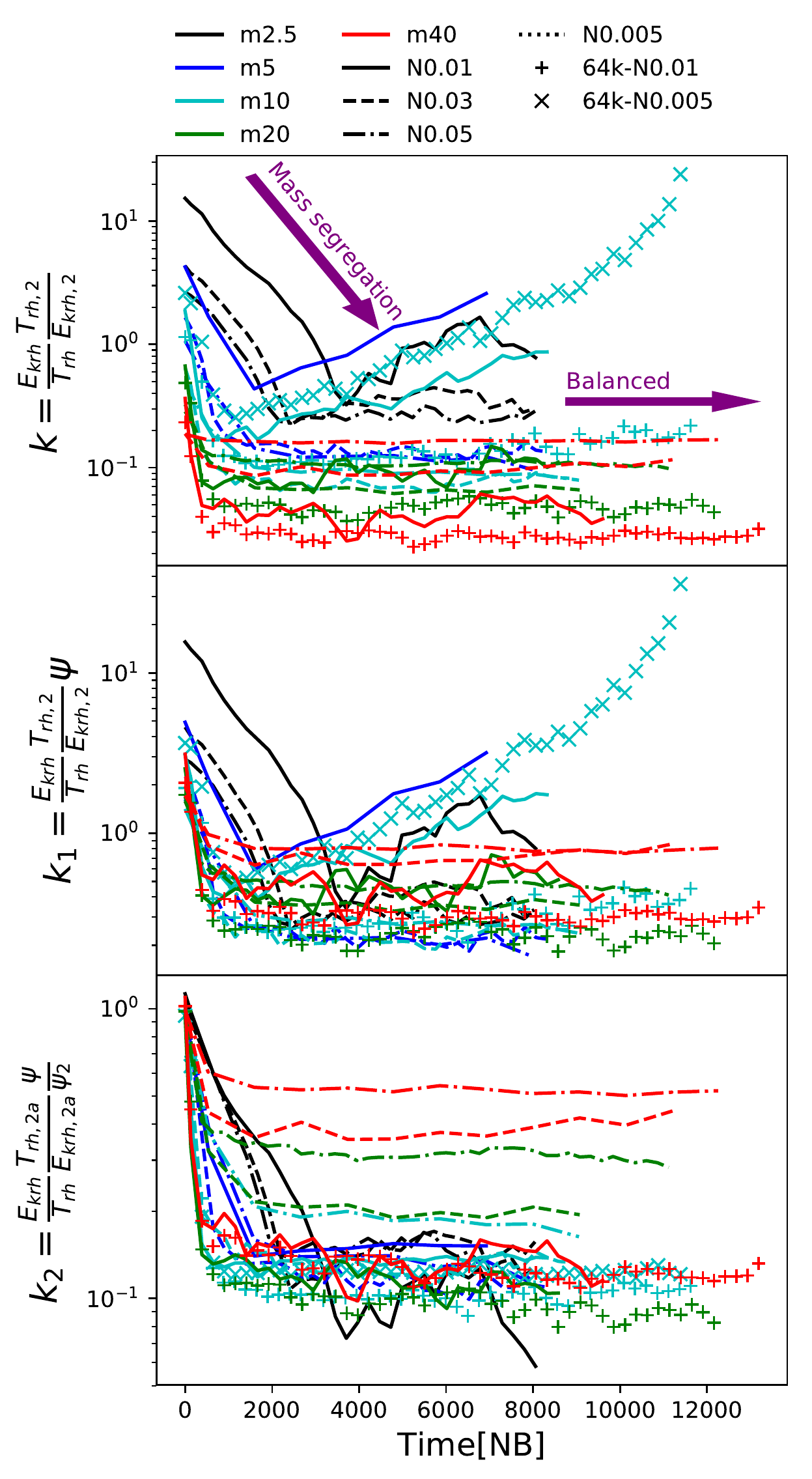}
  \caption{The evolution of the energy-flux ratios between BH subsystems and clusters.
    Upper: $k$ defined in Eq.~\ref{eq:ebk}.
    Middle: the corrected energy-flux ratio, $\ka =  k \psi$ (Eq.~\ref{eq:ebp}).
    Lower: the energy-flux ratio, $\kb$, by considering both two components inside $\rhb$ and the relaxation correction factors, $\psi$ and $\psib$ (Eq.~\ref{eq:ebpp}).
    The time is in the H{\'e}non unit \citep[NB;][]{Henon1971}.
    All models shown here are in an isolated environment.
    Colors represent $\mb/\ms$ (m[value]) and lines indicate $\Nb/\N(0)$ (N[value]).
    The models with initial $64$K particles are shown as $+$ and $\times$ symbols.
    The two purple arrows indicate the two stages of evolution: mass segregation and energy balance.
  }
  \label{fig:erate}
\end{figure}

We first investigate whether the energy-balance theory (Eq.~\ref{eq:eb}) of \cite{Breen2013} applies to star clusters with a large range of $\Mb/M(0)$ in the isolated environment.
Since the four quantities in Eq.~\ref{eq:eb} can be directly measured in the $N$-body simulations, it is easy to calculate the energy-flux ratio, $k$.
To compute the relaxation time, $\Trh$, we use the definition from \cite{Spitzer1987}, 
\begin{equation}
  T_{\mathrm rh} = 0.138 \frac{N^{1/2} \rh^{3/2}}{\m^{1/2} G^{1/2} \ln \Lambda},
  \label{eq:trh}
\end{equation}
where $N$ is the number of objects, $\rh$ is the half-mass radius, $\m$ is the average mass, $\comb$ is the Coulomb logarithm.
We apply $\comb=0.02 N$ based on the measurement of \cite{Giersz1996}.
Instead of the total energy $\E$ and $\Eb$, we use the total kinetic energy $\Ek$ and $\Ekb$ inside $\rh$ and $\rhb$ separately.
Thus, Eq.~$\ref{eq:eb}$ is rewritten as:
\begin{equation}
  \frac{\Ek}{\Trh} \approx k \frac{\Ekb}{\Trhb} .
  \label{eq:ebk}
\end{equation}
This replacement of the energy definition is acceptable when both components are in virial equilibrium.
In Section~\ref{sec:rh}, we show that this assumption is valid for most models.
The result of the energy-flux ratio is shown in the upper panel of Fig.~\ref{fig:erate}.
If the balanced evolution has been achieved, $k$ would keep constant after the mass segregation of BHs.
Indeed, the result shows a two-stage evolution.
At the beginning, $k$ decreases due to the effect of mass segregation.
When the balance is established, most models have a constant $k$ until the end of simulations as expected.
However, a few models with low $\Mb/M(0)$ (e.g. m5-N0.01 and m10-N0.005) show a growth of $k$.
We name them as ``k-growing models''.
Besides, we also see that the absolute values of $k$ at the balanced stage have a large variation depending on $\Nb/N(0)$ and $\mb/\ms$.

To understand the diverged behavior of $k$, we need to consider the two assumptions involved in Eq.~\ref{eq:eb}, \ref{eq:trh} and \ref{eq:ebk}.
Firstly, the relaxation time formula (Eq.~\ref{eq:trh}) is initially derived for a one-component cluster.
We assume that by using the average mass $\m$, $\Trh$ can represent the average relaxation time for multi-component cases.
However, \cite{Spitzer1971} showed that this average is not appropriate. 
It is necessary to introduce a correction factor, $\psi$, to properly calculate multi-component relaxation time \citep[priv. comm. M. Gieles and][]{Antonini2019b},
\begin{equation}
  \Trhp=\frac{\Trh}{\psi} .
  \label{eq:trhp}
\end{equation}
This is because the average diffusion coefficient $\vp$ is not equal to the sum of the coefficients of individual components \citep[Eq. 16 and 17 in ][]{Spitzer1971}.
This results in the correction factor for the multi-component relaxation time, $\psi$, defined by:
\begin{equation}
  \psi = \frac{\sum_k \nk \mk^2/\vk}{\n \m^2 /\vave},
  \label{eq:psi}
\end{equation}
where $\nk$, $\mk$ and $\vk$ are the number density, the mass of one object and the mean velocity of the component $k$. 
$\n$, $\m$ and $\va$ represent the average values of all components respectively.
Here the definition of $\psi$ is different from the original version (Eq.~24) in \cite{Spitzer1971}, where the equipartition of energy is assumed.
Our definition represents a more general condition.
When all quantities are measured within the same $\rh$, $\nk$ can be replaced by the total number of individual components inside $\rh$.

$\psi$ can be directly measured in the $N$-body models.
By including $\psi$ and using $\Trhp$ to replace $\Trh$ in Eq.~\ref{eq:ebk} (shown in the middle panel of Fig.~\ref{fig:erate}),
\begin{equation}
  \frac{\psi \Ek}{\Trh} \approx \ka \frac{\Ekb}{\Trhb} ,
  \label{eq:ebp}
\end{equation}
the new energy-flux ratio, $\ka=k \psi$, has a common value around $0.3$ for most models.
However, k-growing models still exist.

To explain the growth of $\ka$, we investigate another underlining assumption.
\cite{Breen2013} assumes that the cluster center is mostly occupied by BHs, thus the right-hand side of Eq.~\ref{eq:ebk} only includes the BH component.
To obtain a more accurate energy flux inside $\rhb$, we also include the light stars in the calculation of the kinetic energy, $\Ekba$, and the two-component relaxation time with correction factor, $\Trhba/\psib$.
Thus, the new relation is
\begin{equation}
  \frac{\psi \Ek}{\Trh} \approx \kb \frac{\psib \Ekba}{\Trhba},
  \label{eq:ebpp}
\end{equation}
The result is shown in the lower panel of Fig.~\ref{fig:erate}.
Now the new energy-flux ratio has a common stable value, $\kb = 0.1-0.2$, for most cases except the models with $\Mb/\M(0)>0.35$.
When $\Mb/\M(0)=1$, where the system become one-component, based on the definition of Eq.~\ref{eq:ebpp}, the left and right sides are equivalent, i.e., $\Ek \equiv \Ekba$, $\Trh/\psi \equiv \Trhba/\psib$ and $\kb=1.0$.
Thus, it is expected that there exists a transition region of $\Mb/\M(0) \in (\mbmtr, 1.0)$ where $\kb$ is between $0.2$ and $1.0$.
The result in Fig.~\ref{fig:erate} suggests that $\mbmtr \approx 0.35$.
In the next section, we show that this transition criterion is linked to the ratio between the mean velocities of the two components.

\begin{figure}
  \centering
  \includegraphics[width=0.9\columnwidth]{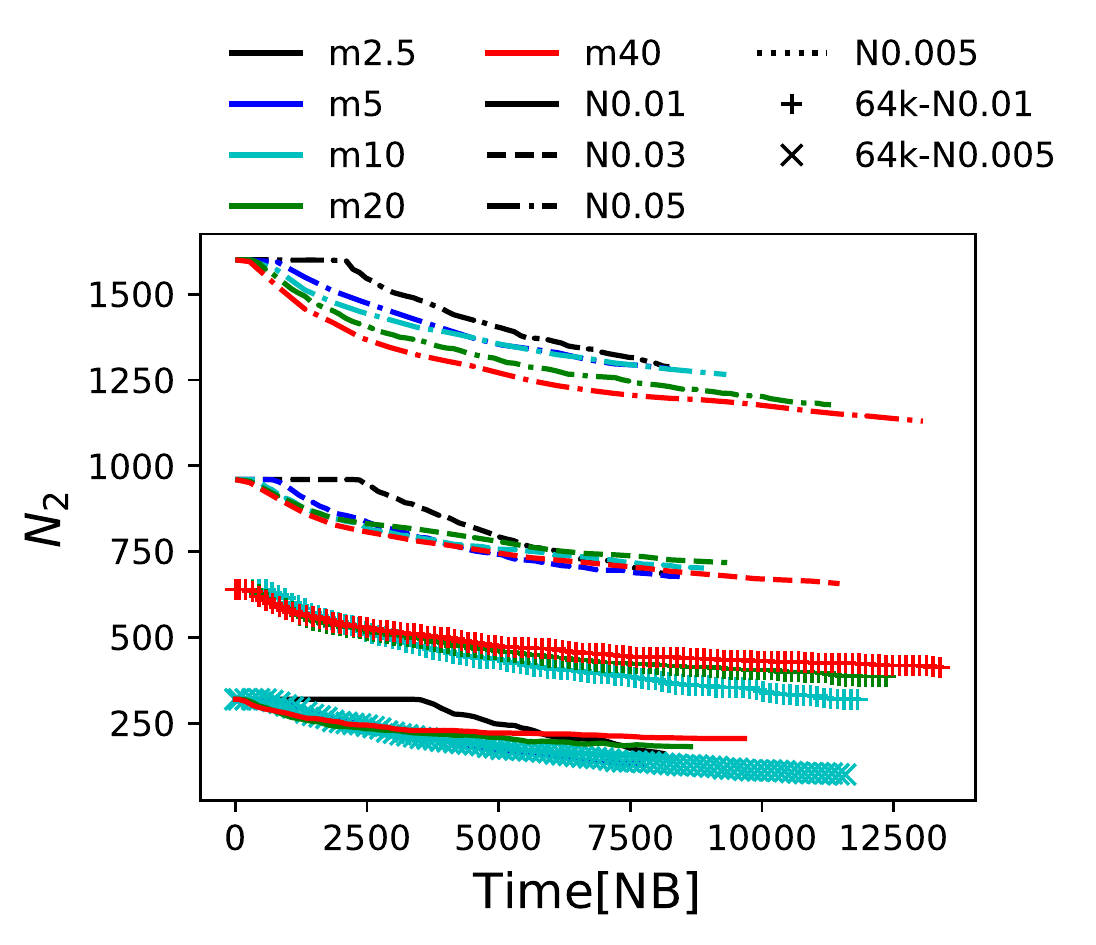}
  \caption{The number of heavy (BH) components, $\Nb$, as a function of time.
    The plotting style is the same as in Fig.~\ref{fig:erate}.
  }
  \label{fig:n2}
\end{figure}

By including both components in the energy flux calculation, the growth trend of $k$ and $\ka$ is absent in $\kb$.
This suggests that the energy flux from light components in the cluster center becomes more and more important in the k-growing models, thus $k$ ($\ka$) increases as BHs become less dominant.
When we consider the escape of BHs, such behaviour can be naturally explained.
In Fig.~\ref{fig:n2}, we show the evolution of the BH number, $\Nb$, for all models.
The k-growing models have a small $\Nb$ initially ($320$).
Once a large fraction of BHs have escaped, the interaction between BHs and light stars in the center becomes important.
On the other hand, Fig.~\ref{fig:erate} shows that the model m10-N0.01 with initially 32k objects (light blue solid curves) has a growth of $k$ while N64-m10-N0.01 (light blue ``$\times$'' symbol) has a stable $k$.
This also suggests that the decreasing number of BHs is the reason for the growth trend of $k$ ($\ka$).

\subsection{Correction factor, $\psi$}
\label{sec:psi}

\begin{figure}
  \centering
  \includegraphics[width=0.85\columnwidth]{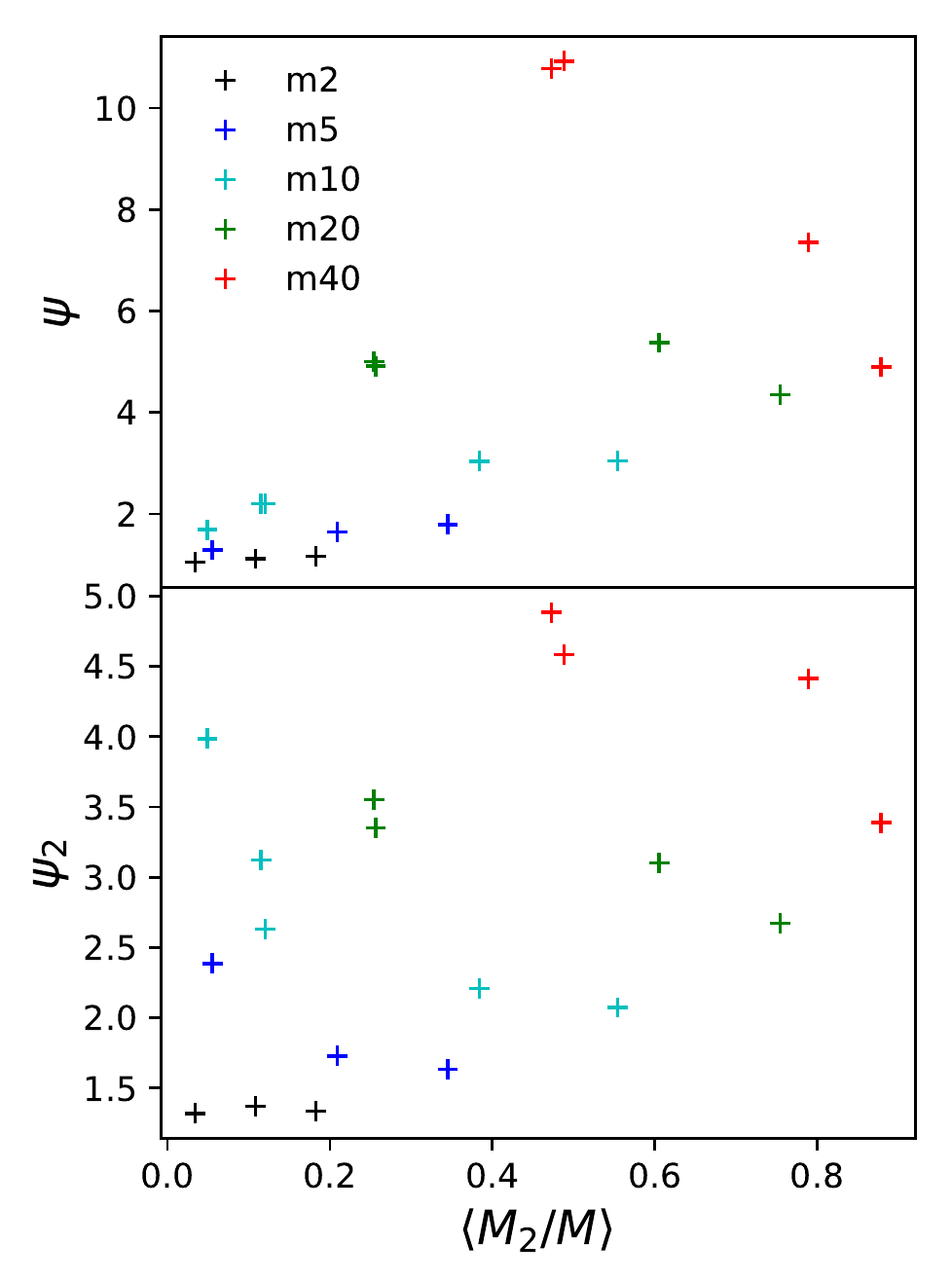}
  \caption{The average correction factor, $\psi$ for $\Trh$ (upper panel) and $\psib$ for $\Trhb$ (lower panel), measured during the balanced evolution.
    The $x$-axis is the average total mass fraction of BH components, $\MbMave$, within $\rh$.
    Colors represent $\mb/\ms$.
  }
  \label{fig:psi}
\end{figure}

It is worth checking the values of $\psi$ and $\psib$, which indicate how large the difference of $\Trhp$ and $\Trh$ can be.
Fig.~\ref{fig:psi} shows the average values of $\psi$ and $\psib$ depending on the average mass fraction of BHs, $\MbMave$, during the balanced evolution.
Notice all quantities for calculating $\psi$ and $\psib$ (Eq.~\ref{eq:psi}) are measured within $\rh$ and $\rhb$ separately.

The variation of $\psi$ and $\psib$ is large.
For the m2.5 models, where the heavy components represent NS like objects, $\psi$ and $\psib$ are close to unity.
When $\mb/\ms$ increases, $\psi$ and $\psib$ also increases significantly.
There is no clear trend how $\psi$ and $\psib$ depends on $\MbMave$ and $\mb/\ms$.
This complexity comes from the additional dependence on the mean velocities of both components (Eq.~\ref{eq:psi}).
When $\mb/\ms\le10$, $\psi$ is positively correlated with $\MbMave$, while when $\mb/\ms\ge20$, there is an anti-correlation.
For $\psib$, except the m2.5 models, all other cases show anti-correlation.


\begin{table}
  \centering
  \caption{The Spitzer mass stratification instability criterion $\chi$ for all models at the initial state. $\chi<0.16$ indicates the Spitzer stable case. Thus except the m2.5-N0.1 model, all other models are Spitzer unstable.}
  \begin{tabular}{llrrr}
    \hline
    $\frac{\mb}{\ms} \backslash \frac{\Nb}{\N}(0)$ & 0.005 &   0.01  &   0.03  &   0.05  \\
    \hline
    2.5 &       &   0.100 &   0.306 &   0.520 \\
    5   &       &   0.565 &   1.729 &   2.942 \\
    10  & 1.589 &   3.194 &   9.780 &  16.644 \\
    20  &       &  18.069 &  55.325 &  94.150 \\
    40  &       & 102.215 & 312.968 & 532.594 \\
    \hline
  \end{tabular}  
  \label{tab:chi}
\end{table}

The relation between the mean velocities of the two components indicate the degree of energy equipartition.
Whether the two-component system can reach complete energy equipartition, $\ms \vs^2 = \mb \vb^2$, depends on $\Mb/\Ms$ and $\mb/\ms$.
The criterion is provided by \cite{Spitzer1987}:
\begin{equation}
  \chi = \Mb/\Ms (\mb/\ms)^{3/2} <\beta, 
  \label{eq:chi}
\end{equation}
where $\beta=0.16$ based on Spitzer's estimation, but may be different depending on the structure of the system.
Table~\ref{tab:chi} show $\chi$ for all models at the initial state.
Except the model m2.5-N0.01, all other models have $\chi>0.16$, i.e., $\ms \vs^2 < \mb \vb^2$ during the balanced evolution.
Thus, without knowing the mean velocities, $\vs$ and $\vb$, it is difficult to estimate $\psi$ only based on $\Mb/M$ and $\mb/\ms$.
However, we can first estimate $\psi$ in two extreme cases.
If $\vb=\vs$, which indicates both components have the same mean velocity, 
\begin{equation}
  \psi = \frac{ \ns \ms^2 + \nb \mb^2}{\n \m^2}.
\end{equation}
In the case of full energy equipartition, $\ms \vs^2 = \mb \vb^2$, 
\begin{equation}
  \psi = \frac{ \ns \ms^{2.5} + \nb \mb^{2.5}}{\n \m^{2.5}}.
\end{equation}
Thus, we can use a power index $\gamma$ to represent the influences from the ratio of mean velocities.
$\psi$ can be written as
\begin{equation}
  \psi = \frac{ \ns \ms^{\gamma} + \nb \mb^{\gamma}}{\n \m^{\gamma}} = \frac{\langle m^{\gamma} \rangle}{\m^{\gamma}}.
  \label{eq:psig}
\end{equation}

\begin{figure}
  \centering
  \includegraphics[width=0.9\columnwidth]{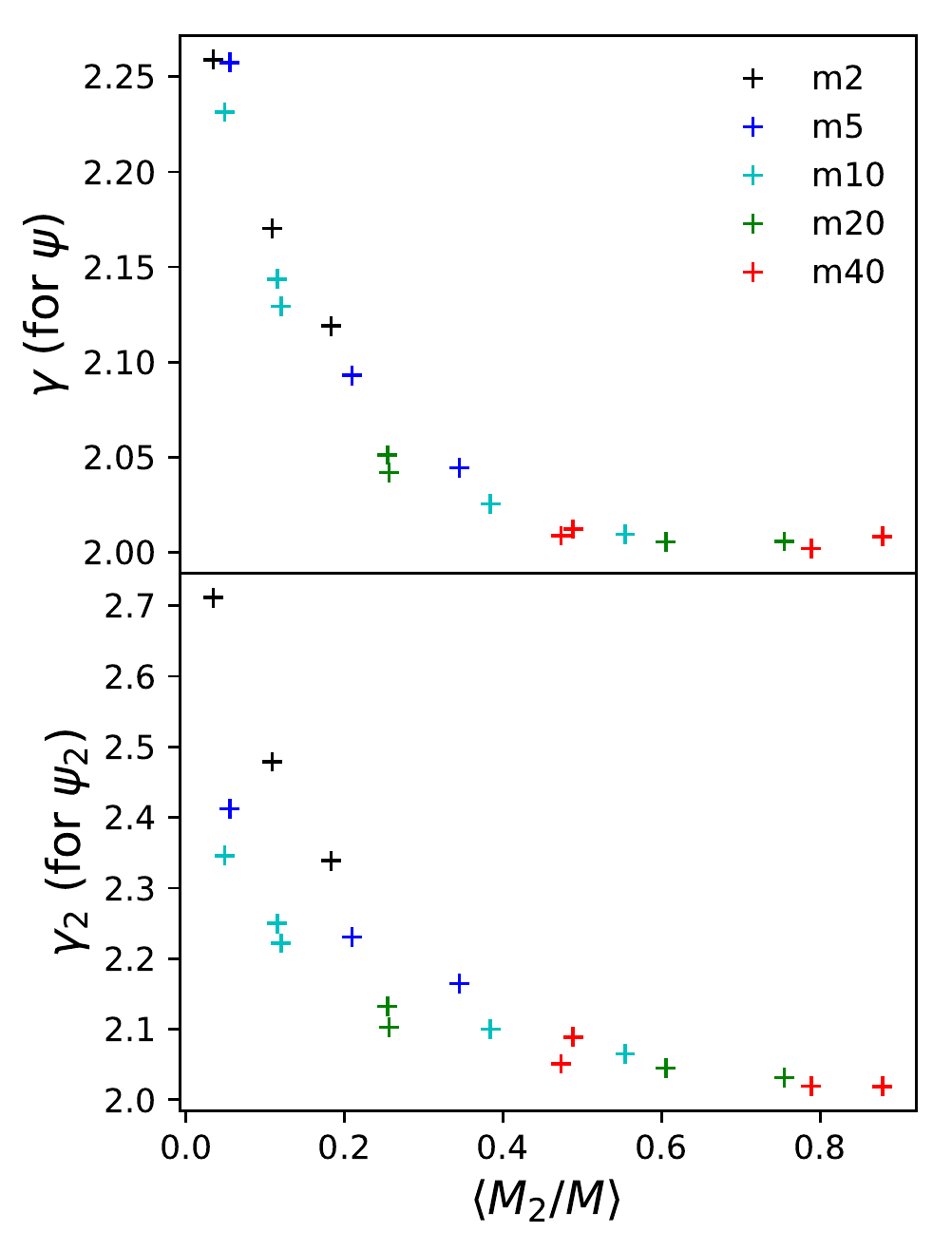}
  \caption{The power index, $\gamma$ for $\psi$ (upper panel) and $\gb$ for $\psib$ (lower panel), from Eq.~\ref{eq:psig}.
    The plotting style is the same as in Fig.~\ref{fig:psi}.
  }
  \label{fig:gamma}
\end{figure}

From the $N$-body models, we measure the $\gamma$ factor for both $\psi$ and $\psib$.
The result is shown in Fig.~\ref{fig:gamma}.
The $\gamma$ of $\psi$ depends on $\MbMave$ but is not very sensitive to $\mb/\ms$.
When $\MbMave$ increases from $0.0$ to $\mbmtr$, $\gamma$ decreases from $2.26$ to $2.0$ and the system departs from energy equipartition.
Then $\gamma$ is stable around $2.0$ and never reach below $2.0$ for $\MbMave>\mbmtr$.
This phenomenon can be understood when the escape of light stars is considered.
The mean escape velocity of the system can be estimated as
\begin{equation}
  \vescm \approx 2 \sqrt{|\phi|},
\end{equation}
where $\phi$ is the total potential of the system, and the mean velocity of the system in virial equilibrium is
\begin{equation}
  \vave \approx \sqrt{|\phi|} .
\end{equation}
Thus
\begin{equation}
  \vescm \approx 2 \vave .
\end{equation}
When $\MbMave>\mbmtr$, heavy components dominate the potential, energy equipartition leads to a high mean velocity of light stars.
Averagely, light stars with velocities larger than $\vescm$ escape the cluster, thus $\vs< \vescm$ should be always satisfied.
On the other hand, $\vb$ becomes closer to $\vave$ when $\MbMave$ increases.
Eventually, $\vs \approx 2 \vb$ in the extreme case, which suggests that $\vb/\vs$ is independent of $\mb/\ms$ and  $\gamma \approx 2.0$.
The transition of $\gamma$ from decrease to flat appears around $\mbmtr$, suggesting that the departure of $\kb$ from $0.1-0.2$ shown in Fig.~\ref{fig:erate} is linked to the enhanced escape of light stars.

The behavior of $\psib$ is similar to $\psi$, but the degree of energy equipartition inside $\rhb$ is higher, i.e. the $\gamma$ of $\psib$ ($\gb$) is larger as shown in Fig.~\ref{fig:gamma}.
$\gb$ also becomes close to $2.0$ when $\MbMave>0.6$.
In the center region, the relaxation and mass segregation times are shorter and the escape velocity (limit for the maximum mean velocity of light stars) is higher compared to the case in the halo.
Thus it is reasonable to see a higher degree of energy equipartition.
$\gb$ shows a stronger dependence on $\mb/\ms$, this explains a larger variation of $\psib$ depending on $\mb/\ms$.


\begin{figure}
  \centering
  \includegraphics[width=0.9\columnwidth]{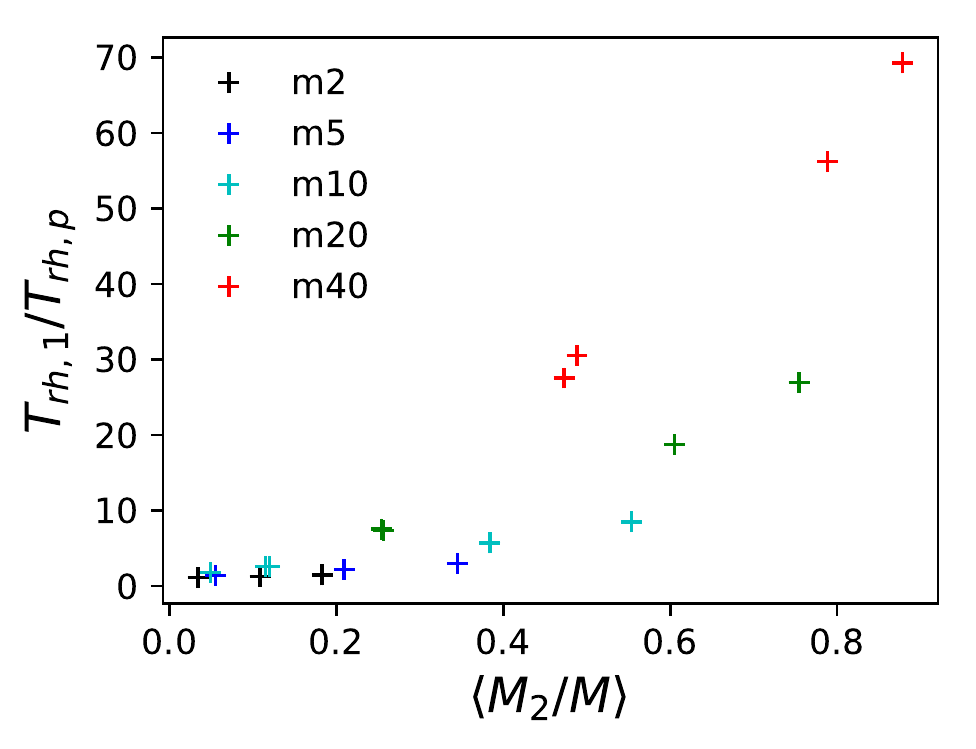}
  \caption{The ratio between the light-component relaxation time, $\Trhs$, and the corrected two-component relaxation time, $\Trhp$. 
    The plotting style is the same as in Fig.~\ref{fig:psi}.
  }
  \label{fig:trhr}
\end{figure}

Observations of star clusters cannot directly measure the two-component relaxation time, $\Trhp$, since BHs are invisible.
Thus we calculate the ratio between light-component relaxation time, $\Trhs$, and $\Trhp$ in Fig.~\ref{fig:trhr}.
This ratio significantly increases with larger $\Mb/M$ and varies in a range of $1-70$.
For $\Mb/M>0.4$, BHs significantly dominate the system, thus the ratio can be above $20$.
In other cases, the ratio is below $10$.
This result indicates that by only measuring the relaxation time of stars, we would significantly overestimate the two-component relaxation time if $\Mb/M$ is large.


%

\subsection{Half-mass radius}
\label{sec:rh}

\begin{figure}
  \centering
  \includegraphics[width=0.9\columnwidth]{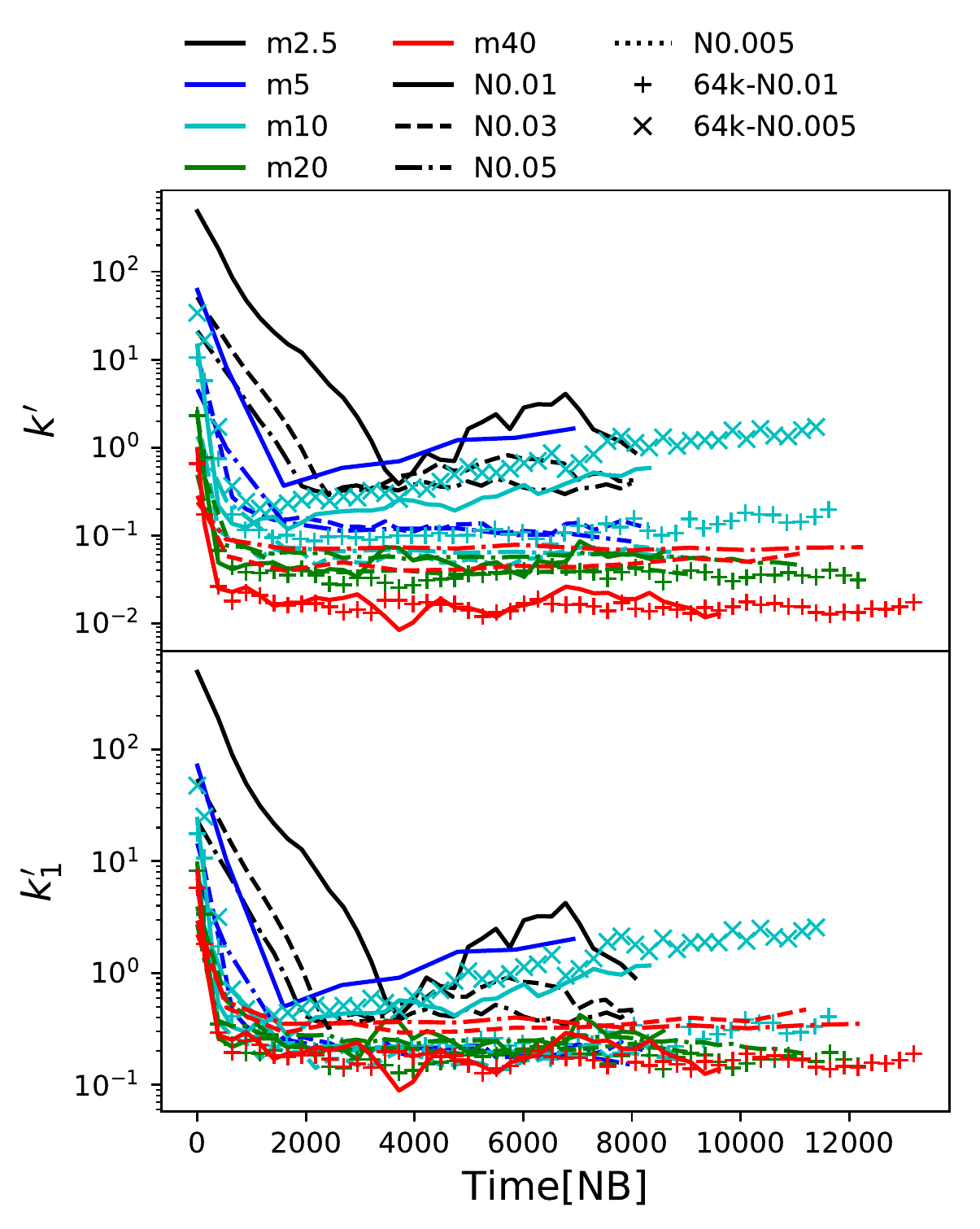}
  \caption{The evolution of $k'$ defined by Eq.~\ref{eq:rhratio} (upper panel).
    The coefficients with correction $\kap=\kp \psi$ are shown in the lower panel.
    The plotting style is the same as in Fig.~\ref{fig:erate}.
  }
  \label{fig:rhratio}
\end{figure}

The importance of $\psi$ should also reflect in the relation of $\rhb/\rh$ shown in the Eq.~4 of \cite{Breen2013}:
\begin{equation}
  \left (\frac{\rhb}{\rh} \right )^{5/2} \approx k' \left (\frac{\Mb}{\M} \right)^{3/2} \frac{\mb}{\m} \frac{\combb}{\comb}.
  \label{eq:rhratio}
\end{equation}
This is derived based on Eq.~\ref{eq:eb}.
When the two components are in virial equilibrium, $k' = k$.

In the upper panel of Fig.~\ref{fig:rhratio}, we compare the coefficient $k'$ of different models defined by Eq.~\ref{eq:rhratio}.
As expected, the differences of $k$ shown in Fig.~\ref{fig:erate} also exist for $k'$.
For the models with flat $k'$ after the balanced evolution start, $k \approx k'$, suggesting that the two components are in virial equilibrium.
The models with increasing $k'$ have $k' < k$ after about $8000$ NB time unit, which suggests a departure from virial equilibrium.

In the lower panel, we apply the correction $\psi$ like the case of Eq.~\ref{eq:ebp}, the differences are reduced, similar to the middle panel of Fig.~\ref{fig:erate}.
Thus, by using Eq.~\ref{eq:psig}, a new relation for the half-mass-radius ratio can be written as
\begin{equation}
  \begin{split}
    \left (\frac{\rhb}{\rh} \right )^{5/2} \approx &
    \kap \left (\frac{\Mb}{\M} \right)^{1/2} \left ( \frac{\mb}{\m} \right )^{2-\gamma} \\
      &  \frac {1}{1 + \frac{\Ms}{\Mb} \left ( \frac{\ms}{\mb} \right )^{\gamma-1}} \frac{\combb}{\comb} ,
  \end{split}
  \label{eq:rhrationew}
\end{equation}
To obtain this relation, the number densities $\ns$, $\nb$ and $\n$ in Eq.~\ref{eq:psig} are replaced by $\Ms/\ms$, $\Mb/\mb$ and $\M/\m$ separately (measured within $\rh$).
Since the corrections of relaxation time and energy flux due to the existence of light stars inside the region of BH subsystems (Eq.~\ref{eq:ebpp}) are not included here, this relation only provides a better but not exactly accurate description of $\rhb/\rh$ compared to Eq.~\ref{eq:rhratio}.
It is valid when $\Mb/M<\mbmtr$ and $\Nb>300$ (a rough boundary).


\subsection{Mass loss}
\label{sec:mloss}

The dissolution time of star clusters depends on the relaxation time.
In this section, we study whether the corrected relaxation time, $\Trhp$, is indeed consistent with the dissolution time of clusters.
In Fig.~\ref{fig:mlossbh} and \ref{fig:mloss},  we compare the evolution of the remaining masses of different components normalized to their initial values, $\Mb(t)/\Mb(0)$ (BHs), $\Ms(t)/\Ms(0)$ (stars) and $\M(t)/\M(0)$ (both), based on different time units.
In each sub-plot, the time is scaled in the way of an integration:
\begin{equation}
  T[\Tscale] = \int^t_{\tst} \frac{dt}{\Tscale(t)},
  \label{eq:tscale}
\end{equation}
where the choice of $\Tscale$ is shown in the $x$-axis labels of sub-plots.
The starting time, $\tst=10~\inttms$, is the roughly measured time of BH core collapse (or the finishing time of BH mass segregation; except for the m2.5 models).
$\Tms$ is the mass-segregation timescale of BHs \citep{Spitzer1987}:
\begin{equation}
  \Tms = \frac{\ms}{\mb}\Trhs.
  \label{eq:tms}
\end{equation}
The normalization factors of the remaining masses, $\Ms(0)$, $\Mb(0)$ and $\M(0)$, are also determined at $\tst$, thus all models have the y-axis value of $1.0$ when $t=\tst$.
This choice of $\tst$ makes a better comparison of the mass loss among different models, because after core collapse, mass loss becomes significant, especially for BHs.
If the choice of $\Tscale$ provides a proper description of the mass-loss timescale, the curves (normalized remaining masses) from different models are expected to overlap with each other.

\subsubsection{BHs}
\label{sec:mlossbh}

\begin{figure*}
  \centering
  \includegraphics[width=2\columnwidth]{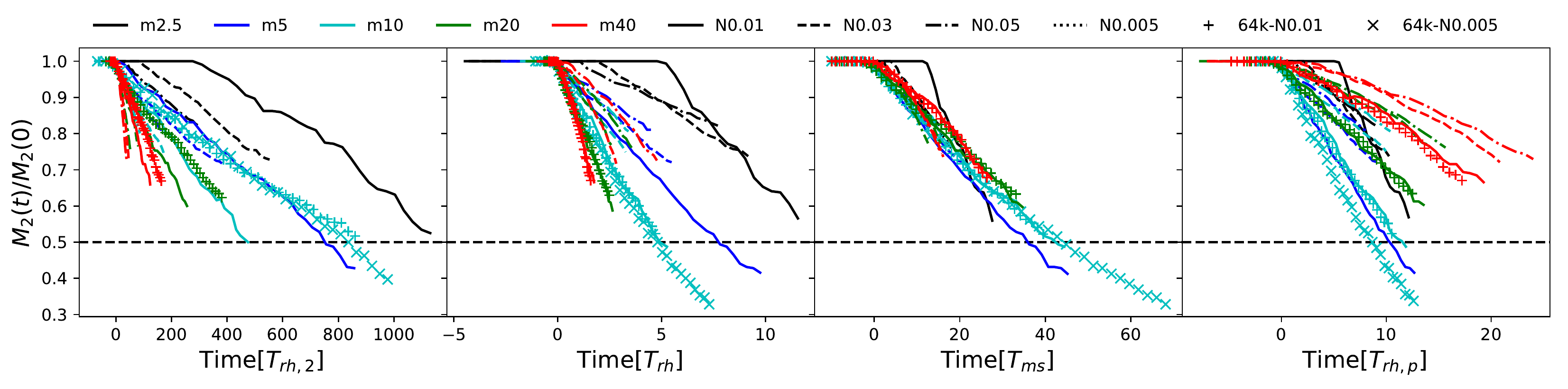}
  \caption{The evolution of the remaining mass of BHs normalized to their initial values.
    From left to right, the four columns show the time in the units of $\Trhb$ (the half-mass relaxation time of BH subsystems), $\Trh$ (the average half-mass relaxation time of clusters), $\Tms = (\ms/\mb) \Trhs$ (the BH mass-segregation timescale) and $\Trhp = \Trh/\psi$ (the corrected two-component half-mass relaxation time) separately (see Eq.~\ref{eq:tscale}).
    For better comparison, $t=0$ is defined at $10~\inttms$ after the starting time of simulations.
    The dashed line indicates $\Mb/\Mb(0)=0.5$.
    The plotting style is the same as in Fig.~\ref{fig:erate}.
  }
  \label{fig:mlossbh}
\end{figure*}

\cite{Breen2013} found that the ejections via few-body interaction should be the major reason for the escape of BHs.
The relaxation driven evaporation mechanism is negligible because most BHs are trapped in the deep potential of the cluster.
They assume that encounters with one hard BH binary in the cluster core release a certain amount of energy before the binary escape the cluster, and this energy generation rate is controlled by the global energy requirement.
On the other hand, each BH binary ejected a fixed number ($5-6$; including the binary itself) of BHs \citep{Goodman1984,Heggie2003}.
Therefore, the escape rate of BHs depends on the global half-mass relaxation time, $\Trh$, and the total mass, $\M$, and is independent on $\Trhb$, as shown in Eq.~9 of \cite{Breen2013}
\begin{equation}
  \Mbdot \sim - M/\Trh .
  \label{eq:bhlosstrh}
\end{equation}
This relation is derived for the case of $\Mb/M\approx0.01$, where the central potential is dominated by the light components.
Consider $\psi>1$ when $\Mb/M$ is large, it is expected that $\Trhp$ should replace $\Trh$, thus
\begin{equation}
  \Mbdot \sim - M \Trhp = - M \psi/\Trh.
  \label{eq:bhlosstrhp}
\end{equation}

Fig.~\ref{fig:mlossbh} show how $\Mb(t)/\Mb(0)$ depends on different scaled time.
Four types of $\Tscale$ are compared: $\Trhb$, $\Trh$, $\Tms$ and $\Trhp$.
The loss of half BHs takes $100-1000$ $\Trhb$ or a few $\Trh$, suggesting that the survival timescale of BHs is much longer than $\Trhb$ and is close to $\Trh$ or $\Trhp$.
This is consistent with the expectation of Eq.~\ref{eq:bhlosstrh} or \ref{eq:bhlosstrhp}.
However, when time is in the units of $\Trh$ or $\Trhp$, the curves $\Mb(t)/\Mb(0)$ show a significant variation of slopes depending on $\Nb/N$ and $\mb/\m$, 
This indicates that there is a missing coefficient that depends on the properties of BH subsystems in Eq.~\ref{eq:bhlosstrh} and \ref{eq:bhlosstrhp}.
Instead, the 3$^{rd}$  panel in Fig.~\ref{fig:mlossbh} shows a well consistent evolution (same slopes) of $\Mb(t)/\Mb(0)$ when time is in the unit of $\Tms$.
The mass loss of BH subsystems can be approximately described by:
\begin{equation}
  \begin{split}
  \Mb(t)/\Mb(\tst) & \approx 0.0125 \int^t_{\tst} \frac{dt}{\Tms} \\
  \end{split}
  \label{eq:bhmloss}
\end{equation}

Now we try to figure out why the expectation of Eq.~\ref{eq:bhlosstrhp} is not exactly reflected in the numerical models.
The idea that one binary eject a fixed number of stars is obtained from the analysis of one-component systems.
In the two-component case, a BH binary ejects both BHs and stars.
In Section~\ref{sec:energy}, we have shown that the light stars are necessary to be included in the calculation of the energy flux in the cluster center.
Thus, the ejection of stars cannot be completely ignored.
Then, the number of ejected BHs per BH binary depends on the number densities of two components in the core, which rely on $\Nb/\N$ and $\mb/\m$.
This dependence is hidden in the missing coefficient of Eq.~\ref{eq:bhlosstrhp}, which causes the diverged behavior of models shown in Fig.~\ref{fig:mlossbh}.

To understand why mass segregation plays the dominant role, we start from a different point of view: the idea of ``loss cone'' from the the studies of supermassive black hole binaries (SMBHBs) in the galactic nuclei \citep[e.g.][]{Lightman1977}.
The loss cone is the region of phase space where stars are on orbits intersecting the orbit of SMBHBs.
These stars are quickly ejected and the loss cone is depleted.
A similar situation can happen in the case of a hard BH binary in the cluster center.
To support the energy requirement, we assume that the core contains at least one hard BH binary in most of the time after core collapse.
If initially the loss cone of the hard BH binary is empty, it can be refilled by BHs suffering orbital changes via perturbations.
These BHs are quickly ejected after an strong interaction with the binary.
In two-component systems, mass segregation is more efficient than two-body relaxation to refill the loss cone.
If a fixed fraction of the ejected BHs exceed the escape velocity of the cluster, it is expected to see that the mass-loss rate of BHs depends on $\Tms$.

\subsubsection{Light stars}

\begin{figure}
  \centering
  \includegraphics[width=\columnwidth]{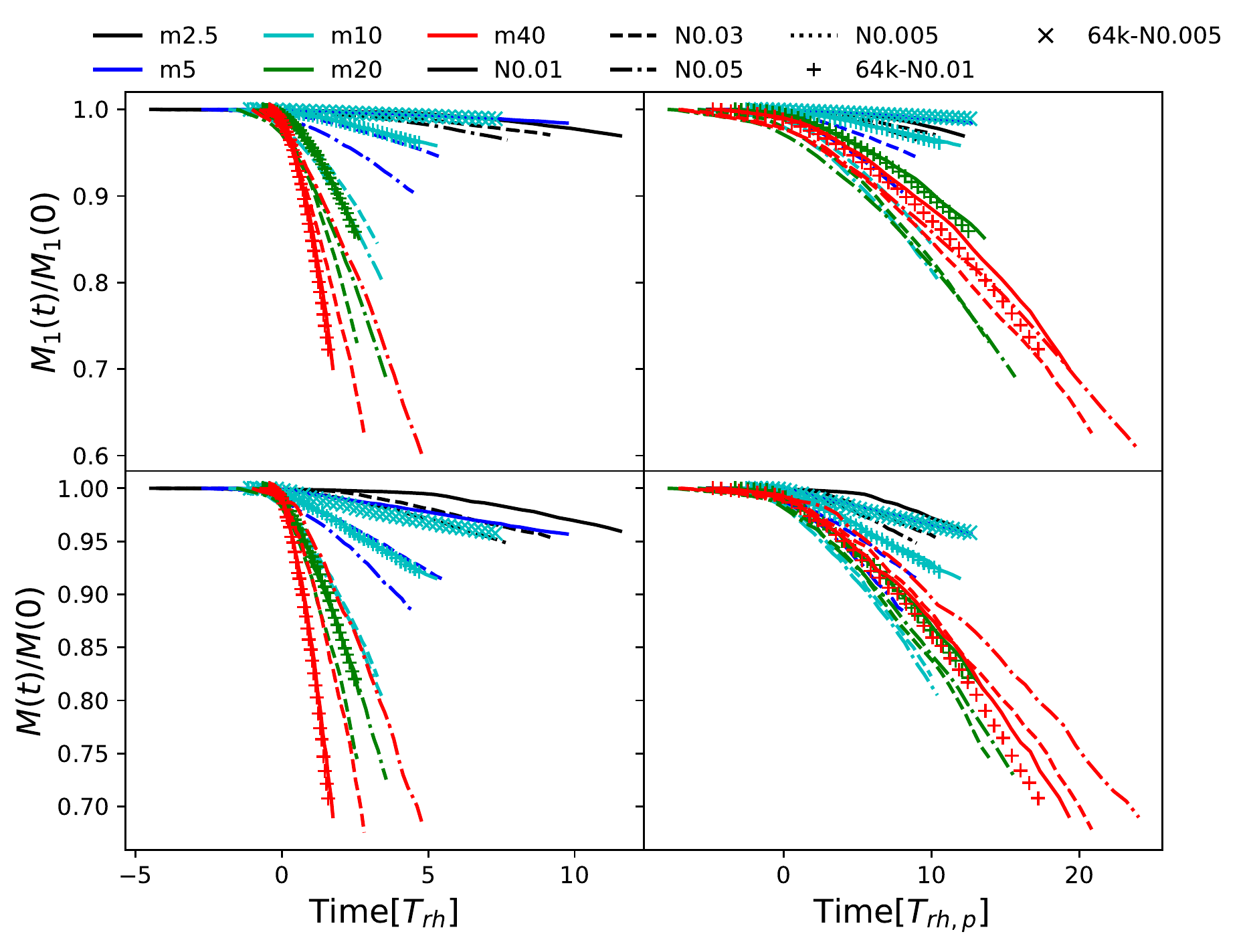}
  \caption{The evolution of the remaining masses of light stars (upper) and both (lower) normalized to their initial values separately.
    The two columns show the time in the units of $\Trh$ and $\Trhp$ separately.
    The plotting style is the same as in Fig.~\ref{fig:mlossbh}.
  }
  \label{fig:mloss}
\end{figure}

In the isolated models, the escape of light stars is driven by the two-body relaxation evaporation and ejections after strong interactions with binaries.
The former is considered as the major channel, thus the mass-loss rate of light stars depends on the relaxation time.
We expect that $\Trhp$ should provide a better match to the mass-loss rate of light stars.
We compare the evolution of the normalized light-component mass, $\Ms(t)/\Ms(0)$, and the normalized total mass, $M(t)/M(0)$, with different scaled times in Fig.~\ref{fig:mloss} (similar like the analysis in Section~\ref{sec:mlossbh}).
The two columns show the times in $\Trh$ and $\Trhp$ separately.
Although a large divergence appears in both cases, $\inttrhp$ indeed provides a better result.

There is also a clear difference for models with $\Mb/M$ below and above $\mbmtr$.
The models of $\Mb/M<\mbmtr$ shows a much slower dissolution.
In Section~\ref{sec:psi}, we show that when $\Mb/M>\mbmtr$, the energy equipartition drives the mean velocity of stars increasing close to the mean escape velocity of the system.
Therefore the escape of light stars is expected to be accelerated when BHs dominate the system.
The speed up of cluster dissolution due to the BH subsystem is also observed in the Monte-Carlo simulations of GCs \citep{Giersz2019}.

\begin{figure}
  \centering
  \includegraphics[width=0.85\columnwidth]{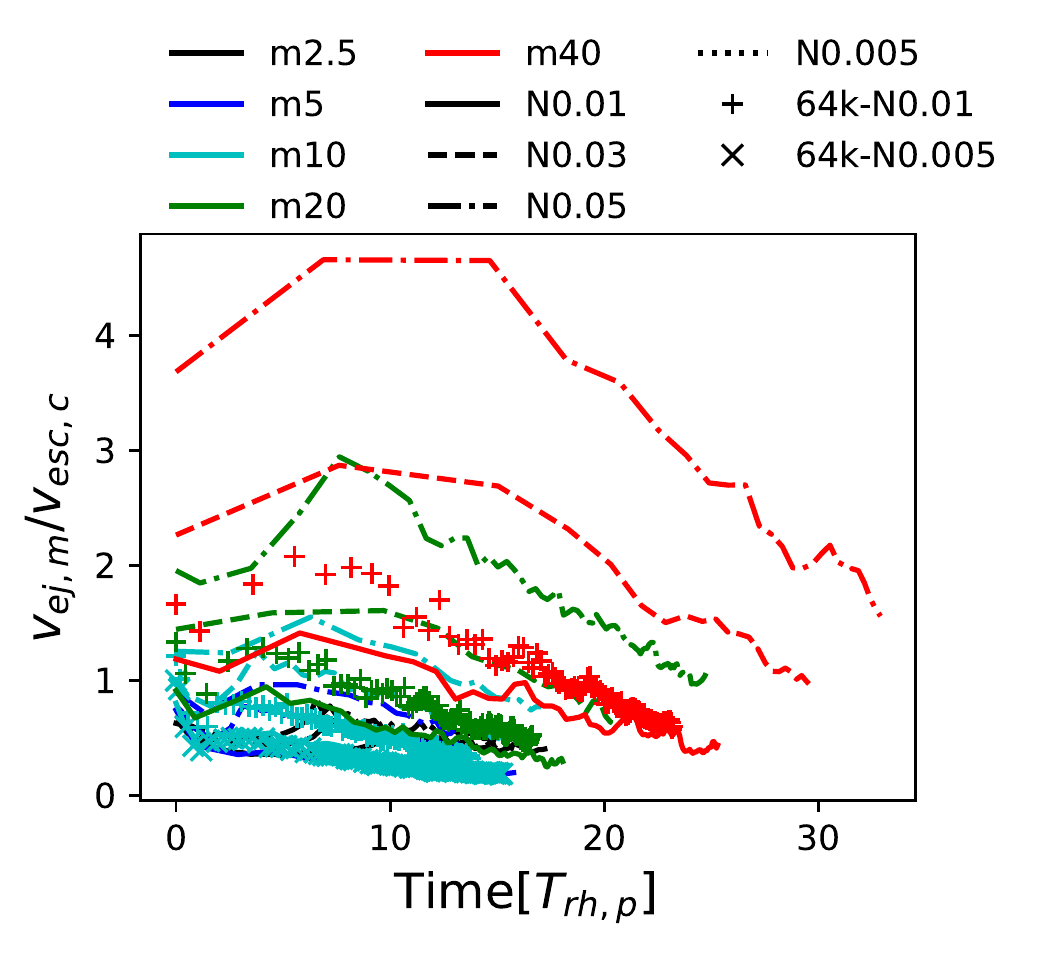}
  \caption{The ratio between the minimum ejection velocity of a star ($\vejm$) from an encounter with a BH binary and the central escape velocity of clusters ($\vesc$).
    The time is in the unit of $\Trhp$.
    The plotting style is the same as in Fig.~\ref{fig:mlossbh}.
  }
  \label{fig:vej}
\end{figure}

On the other hand, we can estimate the ejection velocity of a light star, $\vej$, after an strong interaction with a BH binary in the cluster core by \citep{Miller2009,Antonini2019}:
\begin{equation}
  \vej \approx \left[ 0.2 G \frac{\ms\mb}{2(\ms+2\mb) a} \right]^{1/2} .
  \label{eq:vej}
\end{equation}
where $a$ is the semi-major axis of the binary.
By using $a$ at the half-soft boundary \citep[Heggie-Hills law; ][]{Heggie1975,Hills1975} in BH subsystems,
\begin{equation}
  \ahs = \frac{G \mb^2}{\m \vb^2},
\end{equation}
the lower boundary of $\vejm$ can be estimated.
Considering the mean escape velocity for the Plummer model at the cluster center,
\begin{equation}
  \vesc = \left( 2 \times 1.305 G \frac{M}{\rh} \right)^{1/2},
\end{equation}
we now calculate the ratio, $\vejm/\vesc$, as shown in Fig.~\ref{fig:vej}.
Interestingly, the models with $\Mb/M>\mbmtr$ have $\vejm/\vesc>1.0$.
Thus, it is easier for a star to escape the cluster by one strong binary-single interaction.
This may be another important reason for the fast dissolution.



\section{Tidally filling star clusters}
\label{sec:tidal}

\subsection{Energy flux}
\begin{figure}
  \centering
  \includegraphics[width=0.9\columnwidth]{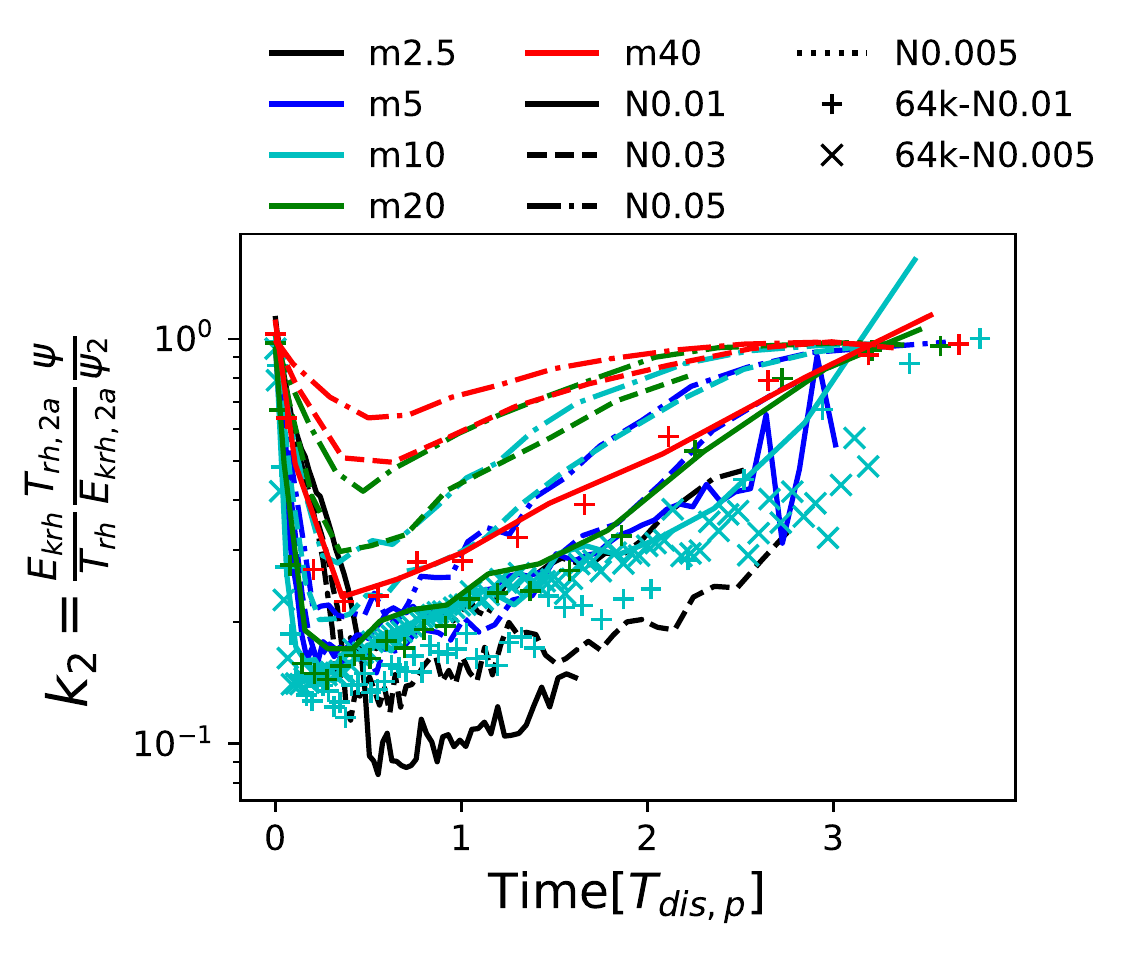}
  \caption{The evolution of the energy-flux ratio for tidally filling models, similar to Fig.~\ref{fig:erate}.
    The time is in the unit of the dissolution time, $\Tdisp(x=1)$, defined in Eq.~\ref{eq:tdistid}.
    }
  \label{fig:eratetid}
\end{figure}

When star clusters are in the galactic tidal field, the tidal stripping becomes an important mechanism to remove stars and BHs from the cluster.
We first investigate how the tidal field influences the balance of energy flux.
Fig.~\ref{fig:eratetid} shows the evolution of energy flux ratio (Eq.~\ref{eq:ebpp}), similar to Fig.~\ref{fig:erate}.
Instead of keeping a constant value as in the isolated star clusters, $\kb$ increases for all models after mass segregation.
For most cases, $\kb$ first decreases to $0.1\sim0.2$ and then increases to about $1.0$ at the end.
The increasing $\kb$ is not necessary to suggest that the energy balance is lost.
Instead, it indicates that the two-component system is smoothly transferred to an one-component system (we also discuss this point in Section~\ref{sec:mbm}).
When a significant fraction of light stars are removed via tidal stripping, $\rhb$ increases towards $\rh$ and the energy flux measured within $\rhb$ does not represent the energy generation at the cluster center any more.

\subsection{Mass loss of BHs}

\begin{figure}
  \centering
  \includegraphics[width=0.8\columnwidth]{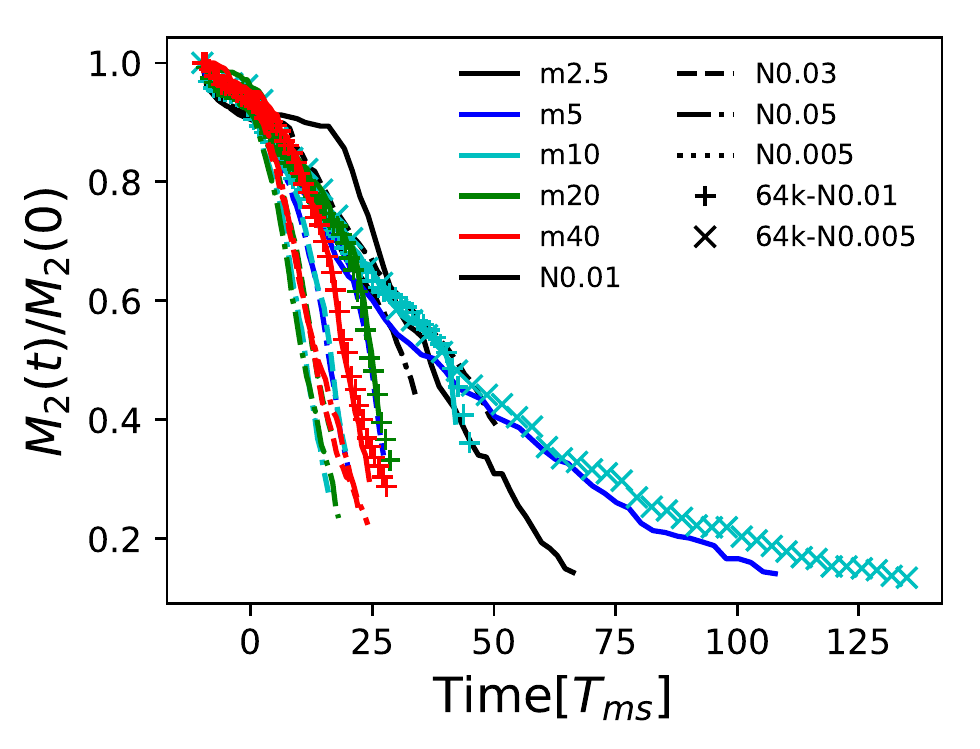}
  \caption{The normalized remaining mass of BHs for tidally filling models depending on the time in the unit of $\Tms$.
    The plotting style is the same as in Fig.~\ref{fig:mlossbh}.
    }
  \label{fig:Mbtid}
\end{figure}

Before the halo of light stars are evaporated, BHs are inside the deep potential.
During this period, the escape rate of BHs is expected to be independent of tidal fields and be similar to the case of isolated clusters, i.e., can be described by Eq.~\ref{eq:bhmloss}.
In Fig.~\ref{fig:Mbtid}, we find that $\Mb(t)/\Mb(0)$ for tidally filling star clusters is still consistent with $\inttms$, but the models with high $\Mb/M$ lose BHs faster.
This is because the high $\Mb/M$ causes faster expansion of $\rhs$, which indicates that light stars are easier to escape.
Eventually when a large fraction of light stars escape, the tidal stripping of BHs becomes more and more important.

\subsection{Dissolution of clusters}

\begin{figure*}
  \centering
  \includegraphics[width=2\columnwidth]{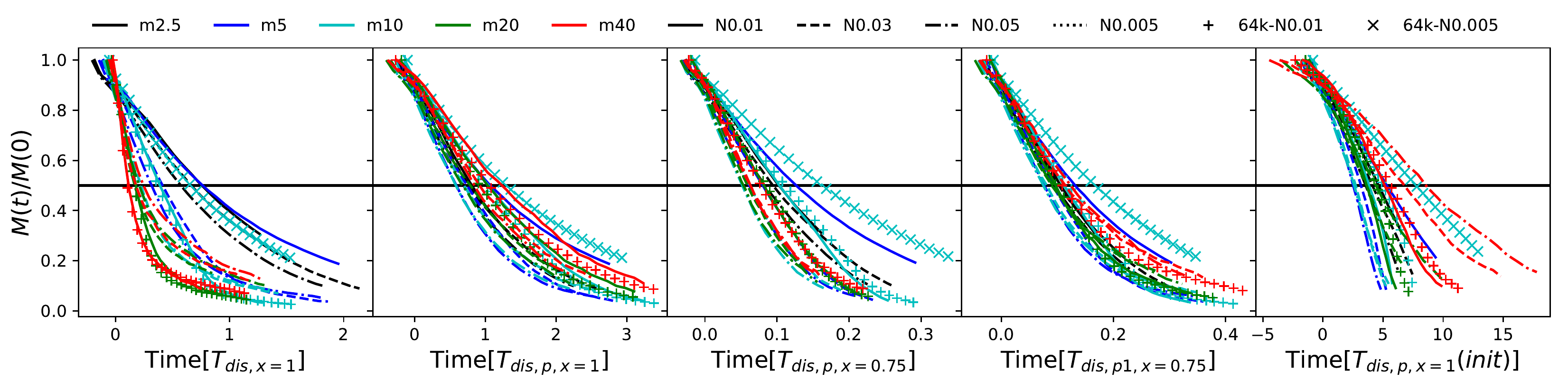}
  \caption{The evolution of total mass normalized to the initial values for tidally filling models.
    The five panels show the time in the units of dissolution times defined as: $\Tdis(x=1)$, $\Tdisp(x=1)$, $\Tdisp(x=0.75)$, $\Tdisps(x=0.75)$ and $\Tdisp(x=1;init)$ separately.
    $\Tdis$ (Eq.~\ref{eq:tdis}) use the average relaxation time, $\Trh$, while $\Tdisp$ and $\Tdisps$ use the corrected value, $\Trhp$.
    $x=1$ indicates no back-scatter effect of potential escapers, while $x=0.75$ indicates the opposite case.
    The last panel shows the time in the unit of the initial value of $\Tdisp(x=1)$.
    The plotting style is the same as in Fig.~\ref{fig:mlossbh}.
  }
  \label{fig:Mtid}
\end{figure*}

The dissolution time of tidal filling star clusters with one-component or Kroupa/Chabrier IMFs are well described by Eq.~\ref{eq:tdis} \citep{Baumgardt2001,Baumgardt2003}.
However, the maximum mass of zero-age main sequence stars in their $N$-body models is $15 M_\odot$.
Thus BH populations are excluded in their analysis.
For the Kroupa/Chabrier IMFs, the final population of BHs is a small fraction of the total mass (see Table~\ref{tab:imf}), the difference between $\Trhp$ and $\Trh$ is small, i.e., the global evolution of the cluster is dominated by the light component.
In our models that represent clusters with top-heavy IMFs, this difference cannot be ignored.
By using $\Trhp$ instead of $\Trh$ in Eq.~\ref{eq:tdis}, the corrected dissolution time is
\begin{equation}
  \Tdisp \sim \Trhp^x\Tcr^{1-x} (\rt/\rh)^{3/2}.
  \label{eq:tdisp}
\end{equation}

But there is one complexity in the two-component models when calculating the radius ratio, $\rt/\rh$, and the crossing time, $\Tcr$.
The latter represents the timescale for a star escaping the system after gaining enough energy to become a potential escaper.
However, the heavy and light components have different half-mass radii and crossing times in the balanced-evolution phase.
Especially when $\Mb/M$ is large, the half-mass radius of stars, $\rhs$, can be much larger than $\rh$ of the system.
Thus, $\Tcr$ calculated by using the averaged property of the system may not correctly represent the escape time of light stars.
To investigate this, we define another dissolution time by using $\rhs$ and $\Tcrs$ (the crossing time measured at $\rhs$):
\begin{equation}
  \Tdisps \sim \Trhp^x\Tcrs^{1-x} (\rt/\rhs)^{3/2}.
  \label{eq:tdisps}
\end{equation}
In Fig.~\ref{fig:Mtid}, we compare the evolution of the total mass, $M(t)/M(0)$,  depending on the time in the units of $\Tdis$, $\Tdisp$ and $\Tdisps$ with $x=1$ and $x=0.75$.
The dissolution time with $x=1$ excludes the back-scatter effect of potential escapers while $x=0.75$ represents the opposite case.
The result shown in Fig.~\ref{fig:Mtid} indicates that $\Tdisp(x=1)$ is the best timescale for describing the cluster dissolution (see the 2$^{nd}$ panel).
By including the back-scatter effect, a large divergence appears among different models, as shown in the 3$^{rd}$ panel.
But once the crossing time and half-mass radius use the values measured at $\rhs$ (the 4$^{th}$ panel), the behaviour of different models becomes much more consistent.
This indicates the importance to have a proper definition of the crossing time and the radius scaling factor in the formula of the dissolution time.
If the time is in the unit of the initial value of $\Tdisp(x=1)$, shown in the 5$^{th}$ panel, the decrease of $M(t)/M(0)$ becomes more linear, but the divergence of different models also become larger.
This is because the dissolution time evolves significantly due to an expansion of the system and a large change of $\Mb/M$.
By using the result of $\Tdisp(x=1)$, we obtain
\begin{equation}
  \M(t)/\M(\tst) \approx 0.5 \int^t_{\tst} \frac{dt}{\Tdisp(x=1)}.
  \label{eq:tdistid}
\end{equation}
This is valid when the cluster is initially tidally filling.


\subsection{The evolution of $\Mb/M$}
\label{sec:mbm}

\begin{figure}
  \centering
  \includegraphics[width=0.9\columnwidth]{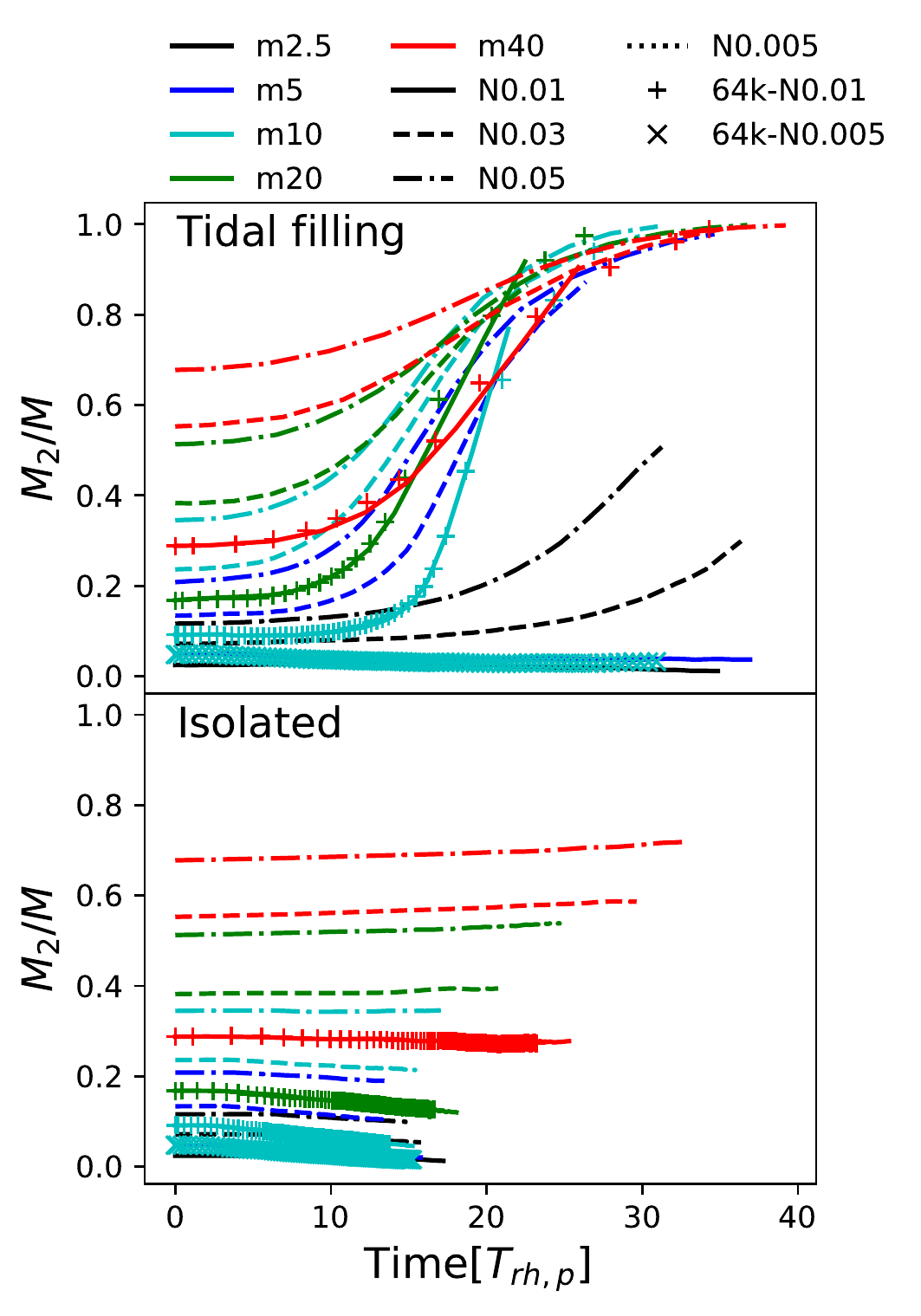}
  \caption{The evolution of $\Mb/M$ for tidally filling and isolated models.
    The plotting style is the same as in Fig.~\ref{fig:mlossbh}.
  }
  \label{fig:mbm}
\end{figure}

Eq.~\ref{eq:bhmloss} and \ref{eq:tdistid} show that the mass loss of BH subsystems and clusters have different timescales.
The BHs escape on the timescale of mass segregation (Fig.~\ref{fig:mlossbh} and \ref{fig:Mbtid}), but the mass loss of light stars depends on the relaxation time (Fig.~\ref{fig:mloss} and \ref{fig:Mtid}).
On the other hand, the dissolution of light components relies on tidal fields, while the escape of BHs does not when they are deeply trapped in the cluster center.
Thus, whether $\Mb/M$ increases or decreases during the long-term evolution is sensitive to the properties of BH subsystems (IMFs) and tidal fields.
Fig.~\ref{fig:mbm} shows that without and with a tidal field, the evolution of $\Mb/M$ is very different.

The isolated models with initial $\Mb/M<\mbmtr$ have a decreasing trend of $\Mb/M$, which indicates that light stars become the dominant component and only few or no BHs stay in clusters after a long-term evolution.
For other models, $\Mb/M$ slowly increases, eventually the systems evolve to dark clusters without light stars.
As discussed in Section~\ref{sec:psi} and \ref{sec:mloss}, when $\Mb/M>\mbmtr$, the escape of light stars is more efficient as a result of energy equipartition.
Thus, it is expected that there is such a transition criterion determining the fate of clusters.

In tidally filling models, due to the tidal stripping of light stars, the evolution of $\Mb/M$ becomes faster.
The transition boundary of $\mbmtr$ also becomes lower.
$\Mb/M$ in the models with initial $\Mb/\M(0)>0.07$ increases fast and finally becomes $1.0$, which means that the systems evolve to BH-dominant dark clusters.
This transition boundary is also discussed in \cite{Breen2013}, where they suggested a constant value that $\mbmtr\approx 0.11$.
The fixed value of $0.11$ is obtained based on the assumption that the mass loss of light stars and BHs both depend on the relaxation time and the total mass.
Although $0.11$ is not far from the value in our result, the boundary, $\mbmtr$, is probably not a constant but depends on $\mb/\ms$, $\Mb/M$ and $\rt/\rh$.
Fig.~\ref{fig:mbm} shows that the slope of $\Mb/\M$ for different models is not the same, which reflects such dependence.
On the other hand, the result in Fig.~\ref{fig:mbm} indicates that when a star cluster has an IMF with $\aimf>-2.0$, where $\Mb/M(0) \ge 0.182$, it can easily evolve to a dark cluster after about $20~\Trhp$ in a strong tidally filling environment.

\section{Discussion and conclusion}
\label{sec:conclusion}

In this work, we have carried out a series of $N$-body simulations of star clusters with two components (BHs and light stars), which approximates star clusters with top-heavy IMFs.
The simplification of the $N$-body models allows us to focus on the dynamical effect of BH subsystems in the long-term evolution of star clusters.
We validate the idea of energy-flux balanced evolution based on the \cite{Henon1975} principle and \cite{Breen2013}.
The result (Fig.~\ref{fig:erate}) indicates that to properly calculate the two-body relaxation time of two-component clusters, a correction factor $\psi$ to the Spitzer relaxation time for average cluster properties, $\Trh$ (Eq.~\ref{eq:trh}), is necessary ($\Trhp=\Trh/\psi$).
Our models show that $\psi = 1-10$ for star clusters with the total mass fraction of BHs, $\Mb/M$, in the range of $2.5\%-68\%$.
Since $\psi$ depends on the property of BHs (Eq.~\ref{eq:psi} and \ref{eq:psig}), the evolution of star clusters, especially the dissolution time (Eq.~\ref{eq:tdisp} and \ref{eq:tdisps}), are sensitive to $\Mb/\M$ and the mass ratio, $\mb/\ms$.
Thus, star clusters with top-heavy IMFs tend to dissolve faster.
For example, with $\aimf=-1.7$ (see Table~\ref{tab:init}), the initial $\Mb/\M\approx46.6\%$, the corresponding $\psi\approx10$.
It is expected that such clusters can dissolve $10$ times faster than the estimation based on the relaxation time for average cluster properties.
Moreover, in observations BHs cannot be directly detected, thus the measurement of relaxation time only counts light stars.
The comparison of $\Trhp$ and the light-component relaxation time, $\Trhs$, shows that $\Trhs$ can be $1-70$ times larger than $\Trhp$.
This suggests that if we ignore BHs in a star cluster, the relaxation time of the system can be significantly overestimated.

The $\psi$ factor can be simplified as Eq.~\ref{eq:psig} with a power index $\gamma$.
The value of $\gamma$ represents the degree of mass segregation of two-component systems.
In Fig.~\ref{fig:psi} and \ref{fig:gamma}, we obtain the average $\psi$ and $\gamma$ depending on $\mb/\ms$ and $\MbMave$.
The average is calculated by using data from the start of the balanced evolution to the end of simulations.
For models with small values of $\MbMave$ which represent the case of the canonical IMF, the system has a high degree of mass segregation with $\gamma$ being close to $2.5$.
When $\Mb/M>0.35$ (top-heavy IMF), the mean velocities of two components tend to be close to each other and be independent of $\mb/\ms$ ($\gamma \approx 2.0$), because the mean velocity of light stars is limited by the mean escape velocity of the system.

The escape rate of BHs depends on the mass segregation time (Eq.~\ref{eq:bhmloss}, Fig.~\ref{fig:mloss},\ref{fig:Mbtid}) and is not sensitive to the tidal field (when light stars still dominate the halo of clusters).
This is different from the mass-loss rate of light stars (Fig.~\ref{fig:mloss} and \ref{fig:Mtid}).
In the isolated cluster, $\Trhp$ is a better timescale than $\Trh$ to describe the mass loss of light stars.
However, the dissolution time shows a very diverged behavior between clusters with low and high $\Mb/M$.
In the case of high $\Mb/M$, light stars are easier to reach the escape criterion due to the energy equipartition and strong interactions with BH binaries.
In the tidal filling case, the dissolution time of the system can be well described by $\Trhp$ (Eq.~\ref{eq:tdisp}) without the back-scatter effect of potential escaper ($x=1$) or with the effect but only if the crossing time and radius scaling factor is properly chosen (Eq.~\ref{eq:tdisps}).

Because the mass-loss rate of BHs and light stars depends on different timescales, the initial $\Mb/M$ and the strength of tidal fields determine the fate of star clusters.
Roughly, with a top-heavy IMF that $\aimf<-2.0$ and a weak tidal field, they eventually evolve to ``luminous clusters'' where very few BHs exist.
With $\aimf>-2.0$ or a strong tidal field, ``dark clusters'' dominated by BHs easily form.
For isolated star clusters, the boundary of $\MbMave$ for this transition is around $\mbmtr=0.3-0.4$.
For tidally filling clusters, it is about $0.07$, which is the case of the canonical IMF.
The timescale to become a dark cluster is about $20~\Trhp$ in a strong tidal filling environment.
This suggests that if massive star clusters (e.g. GCs) with top-heavy IMFs have formed at the high redshift and have been close to the Galactic center, they were easier to be evaporated by tidal fields or have become ``dark clusters'' that cannot be observed by optical telescopes today.
On the other hand, the post-core collapse GCs with dense cores \citep[e.g. $47$~Tuc;][]{Henault-Brunet2019} do not tend to have top-heavy IMFs (unless most BHs have escaped after supernovae with high-velocity natal kicks).
The future observations of gravitational lensing and gravitational waves may provide a constraint on the population of dark clusters.

In this work, we only theoretically focus on the dynamical effect of BH subsystems for the long-term evolution of two-component, spherical and low-$N$ star clusters.
These simple models allow us to well isolate the dynamical effect of BH subsystems from other physical processes.
In a realistic situation, primordial binaries, stellar evolution, mass spectrum, evolution of tidal field, density profiles and rotation of clusters complicate the simple picture discussed here \citep[e.g.][]{Giersz2019}.

Especially, the mass spectrum of stars and BHs introduces an additional complexity in determining the proper relaxation time.
We assume that the equal mass of BHs (stars) can reasonably represents the averaged properties of the objects with a mass spectrum.
However, the importance of $\psi$ discussed in this work suggests that such approximation may not be accurate when the mass spread is wide.
From Fig.~\ref{fig:psi}, we see that when $\mb/\ms>5$, $\psi$ becomes larger than $1.0$.
In an evolved star cluster, the mass spectrum of light stars can ranges from $<0.08$ (brwon dwarfs) to $\sim 1.0$~M$_\odot$, thus $\psi$ for only light stars may already be larger than $1.0$.
In such case, the calculation of an accurate relaxation time requires an integration of the diffusion coefficients of different mass components.
However, \cite{Baumgardt2003} indicated that a star cluster with a mass spectrum but no BHs can be well described by Eq.~\ref{eq:tdis} without the correction factor ($\psi$).
This suggests that compared to the strong impact of BH subsystems shown in this work, the effect introduced by the mass spectrum of light stars is not significant. 
Thus the two-component system should provide a reasonable representation of a star cluster with a top-heavy IMF.

On the other hand, the stellar evolution brings another complexity.
Especially, the stellar evolution models have large uncertainty, how does the kick velocity distribution of BHs and NSs after supernovae looks like is still an open question.
But this distribution is crucial to determine the remaining total mass fraction of BHs and NSs.
This leads to an uncertainty to map an IMF to the initial values of $\Mb/M$ and $\mb/\ms$.
Besides, $\mb/\ms$ also depends on the metallicity.
Thus, our results only provide an approximate analysis to show how to connect the shape of IMFs to the dynamical evolution of star clusters.

The existence of primordial binaries provides an additional energy source and increases the possibility of binary-binary interactions.
It may influence the core radius of clusters and the energy generation.
However, we expect the global evolution (mass loss and energy balance) may not be significantly influenced because whatever the source of energy generation looks like, the energy generation rate should be always guided by the global energy requirement (H{\'e}non's principle).

We focus on the long-term evolution of star clusters.
In the formation stage of clusters embedded in the gas cloud, the influence of stellar-wind feedback from massive OB stars to the gas can be strongly enhanced when IMFs become top-heavy.
Thus, whether the star formation region with a top-heavy IMF can become a gravitational bound cluster or an association after gas is removed is an open question.
In the follow-up projects, it is necessary to validate our conclusion by including these multiple physical effects step by step.

\section*{Acknowledgments}
L.W. thanks the Alexander von Humboldt Foundation for funding this research.
L.W. also thanks Douglas C. Heggie, Mark Gieles, Mirek Giersz, Anna Lisa Varri and Sverre Aarseth for very useful discussions and advice for this project.


\label{lastpage}

\end{document}